\documentclass[11pt]{article}

\usepackage[english]{babel}
\usepackage{amssymb}
\usepackage[letterpaper,top=2cm,bottom=2cm,left=3cm,right=3cm,marginparwidth=1.75cm]{geometry}

\usepackage{amsmath}
\usepackage{graphicx}
\usepackage[colorlinks=true, allcolors=blue]{hyperref}

\title{\textbf{Adiabatic Inspiral Transition and Induction to Plunge of a Compact Body in Equatorial Plane Around a Massive Kerr Black Hole} }
\author{Boyan Wang}
\date{May 2024}

\begin{document}
\maketitle

\begin{abstract}
This paper reconstructs the derivation process from the Kerr Metric to the adiabatic inspiral, transition, and plunge regimes, aiming to meticulously points out the details and logic that were overlooked or skipped in the derivation process. The first half of the paper aims to provide readers with a complete road-map, enabling those with understanding of advanced general relativity to grasp the full logic of the adiabatic inspiral/transition/plunge regime derivation solely through this paper. The latter half addresses the disjointed issue from the Adiabatic Inspiral to the Plunge, including analyses and reinterpretations of the Ori-Thorne transition procedure and the Ori-Thorne-Kesden transition procedure, and proposes two new interpretations: a variant form of the Kesden Y Correction and the Adiabatic Inspiral Perturbation Induced Plunge. The paper also introduces the concept of the Most Stable Circular Orbit(MSCO) during the derivation, and analyses the properties of this Characteristic Radii of Kerr Black Holes. 
\end{abstract}

\section{Introduction}

The success of the Laser Interferometer Gravitational-Wave Observatory (LIGO) project has significantly fueled interest and hope for the Laser Interferometer Space Antenna (LISA) project. LIGO's direct detection of gravitational waves in 2015 confirmed a major prediction of Einstein's theory of General Relativity, opening up a new way of observing the universe. This success acts as a proof of concept for LISA, which aims to detect gravitational waves in a lower frequency band, from sources not accessible to LIGO\cite{Abbott_2009}\cite{amaroseoane2017laser}. 

One of the most exciting possible observations from the Laser Interferometer Space Antenna (LISA) involves the phenomena known as adiabatic inspiral and plunge. This process describes the gradual spiral motion of compact objects, such as black holes or neutron stars, as they orbit each other and lose energy in the form of gravitational waves. As the objects draw closer, the rate of energy loss increases, leading to a faster inward spiral. Eventually, the objects reach a critical separation at which the inspiral phase ends, and the plunge phase begins. During the plunge, the objects rapidly coalesce into a single entity, merging in a highly dynamic and energetic event. This transition from inspiral to plunge is a critical phase in the life cycle of binary systems and represents a key target for gravitational wave observatories like LISA.

LISA's sensitivity to low-frequency gravitational waves makes it ideally suited to observe these inspiral and plunge events, especially from massive black hole binaries in the centers of galaxies. By detecting these processes, LISA could provide invaluable insights into the nature of compact objects, the dynamics of binary systems, and the fundamental physics of gravity. Additionally, studying these events would allow scientists to test General Relativity in extreme conditions, explore the formation and growth of supermassive black holes, and understand more about the structure and evolution of galaxies.

There have been numerous attempts to solve Kerr geodesics; however, solving the full Kerr geodesics for intricate and evolving systems such as extreme mass ratio inspirals (EMRIs) presents formidable challenges. These include computational intensity and the inherent complexity arising from the non-linear nature of Kerr geodesics.  

In light of these challenges, the adiabatic approximation emerges as a vital tool. It greatly simplifies our calculations by allowing us to treat the orbital evolution as a slow, continuous process, in stark contrast to the rapid, dynamic changes of the complete system. This approach effectively decouples the fast orbital motion from the slower inspiral process. Consequently, we intend to explore how the trajectory of a massive test particle can be approximated under the adiabatic inspiral and further extended framework, thereby offering a more tractable and yet physically insightful understanding of these complex astrophysical phenomena.

\section{Kerr Metric}
In our discussion we will use Kerr metric \cite{PhysRevLett.11.237}(with metric signature (-,+,+,+) under Boyes-Lindquist \cite{1967JMP.....8..265B} coordinate as:

\begin{align*}
ds^2 = &-\left(1-\frac{2Mr}{\Sigma}\right) dt^2 - \frac{4Mar\sin^2\theta}{\Sigma} dt d\phi + \frac{\Sigma}{\Delta} dr^2 \\
&+ \Sigma d\theta^2 + \left(r^2+a^2+\frac{2Ma^2r\sin^2\theta}{\Sigma}\right)\sin^2\theta d\phi^2, \\
\end{align*}
or equivalently,
\begin{align*}
g_{tt} &= -\left(1 - \frac{2Mr}{\Sigma}\right), \\
g_{t\phi} &= g_{\phi t} = -\frac{2Mar\sin^2\theta}{\Sigma}, \\
g_{rr} &= \frac{\Sigma}{\Delta}, \\
g_{\theta\theta} &= \Sigma, \\
g_{\phi\phi} &=  \left(r^2 + a^2 + \frac{2Ma^2r\sin^2\theta}{\Sigma}\right)\sin^2\theta,
\end{align*}
$$\Delta=r^2-2Mr+a^2,\Sigma=r^2+a^2cos^2\theta.$$
Here $M$ is the mass of the black hole, $a$ is its angular momentum ($J$) per unit mass, $a:=J/M$ cf. \cite{1972ApJ...178..347B}
\section{Fundamental governing equations}
In this chapter, we will derive the geodesic equations governing the tr,Eajectory of a massive particle orbiting in the equatorial plane of a rotating black hole. Our initial focus will be on examining the time and azimuthal coordinates equations. Subsequently, we apply the simplifications provided by the equatorial plane to derive the radial part of the equation. Finally, we will assemble all the fundamental governing equations necessary for our study.

\subsection{Time-Like Kerr Geodesics}

In the process of getting Kerr geodesics, we use geometrised units for gravitational constant and speed of light, i.e. G=c=1, and we introduce Lagrangian $\text{}\mathcal{L} = \frac{m}{2}g_{\mu\nu}\dot{x}^\mu \dot{x}^\nu$, and thus obtained
\begin{align*}\mathcal{L} = \frac{m}{2} \left[ -\left(1 - \frac{2Mr}{\Sigma}\right) \dot{t}^2 - \frac{4Mar\sin^2\theta}{\Sigma} \dot{t} \dot{\phi} + \frac{\Sigma}{\Delta} \dot{r}^2 + \Sigma \dot{\theta}^2 + \left(r^2 + a^2 + \frac{2Ma^2r\sin^2\theta}{\Sigma}\right)\sin^2\theta \dot{\phi}^2 \right],\end{align*}
which $\text{}\dot{x}=\frac{dx}{d\tau}$.Noticing the $t$ and $\phi$ independence from Lagrangian, and using E-L equation:
\begin{align*}\dot{p_t}:=\frac{\text{d}}{\text{d}\tau}(\frac{\partial \mathcal{L}}{\partial \dot{t}})=\frac{\partial \mathcal{L}}{\partial t}=0,\:\:\:\dot{p_\phi}:=\frac{\text{d}}{\text{d}\tau}(\frac{\partial \mathcal{L}}{\partial \dot{\phi}})=\frac{\partial \mathcal{L}}{\partial \phi}=0,\end{align*}
implies:
\begin{align*}
p_t&=\frac{\partial \mathcal{L}}{\partial \dot{t}}=\frac{m}{2}[\dfrac{\left(4Mr-2{\Sigma}\right)}{{\Sigma}}\dot{t}-\dfrac{4Mar\sin^2\left({\theta}\right)}{{\Sigma}}\dot{\phi}]=constant=-E,\\p_\phi&=\frac{\partial \mathcal{L}}{\partial \dot{\phi}}=\frac{m}{2}[-\dfrac{4Mar\sin^2\left({\theta}\right)}{{\Sigma}}\dot{t}+2\left(\dfrac{r^2+a^2+2Ma^2r\sin^2\left({\theta}\right)}{{\Sigma}}\right)\sin^2\left({\theta}\right)\dot{\phi}]=constant=L.
\end{align*}
Where $E$ represents the energy of the particle, and $L$ denotes its angular momentum. Both of them are constants. Throughout this article, we will frequently refer to $E$ and $L$. It's noteworthy that some papers adopt a slightly modified approach when formulating equations involving $E$ and $L$, or they might assign to these variables signs opposite to those used here. However, such distinctions are not critical for our analysis. This is because in general relativity, particularly in curved spacetime like the Kerr metric, the concept of energy is more nuanced than in Newtonian physics. Therefore, the differences in defining energy and angular momentum is do not significantly impact our discussion. 

Then, we use $\begin{cases}p_t(\dot{t},\dot{\phi})=-E \\p_\phi(\dot{t},\dot{\phi})=L &\end{cases}$ to solve for $\dot{t}$ and $\dot{\phi}$ :
\begin{equation}
\dot{t}=-\frac{2arM}{m\Sigma\Delta}L+\frac{1}{m\Delta\Sigma}[\Sigma(a^2+r^2)+2Ma^2r\sin^2\left({\theta}\right)]E,
\label{eq:t geodesic equation} 
\end{equation}
\begin{equation}
    \dot{\phi}=\frac{\Sigma-2Mr}{m\Delta\Sigma\sin^2\left({\theta}\right)}L+\frac{2arM}{m\Delta\Sigma}E.
\label{eq:phi geodesic equation} 
\end{equation}
For the geodesics in radial and polar coordinates, we will initially employ certain simplifications. The complete solutions for time-like geodesics are detailed in \cite{1972ApJ...178..347B} [2.9]. From these solutions, it is evident that the geodesic equations for both azimuthal and time coordinates align with our findings. 

\subsection{Simplification: equatorial plane}
We have found two governing equations for time and azimuthal coordinate, and now we need to find similar equations for radial and polar coordinate. There are several ways to look at this, and since as we are interested in massive particle in equatorial plane, we are going to use these simplification by forcing $\theta=\pi/2$ , $p_{\theta}=\dot{\theta}=0$ . Thus we have $\sin^2(\theta)=1,\cos(\theta)=0,\Sigma=r^2$.
Now we use the fact that time-like particles follows \begin{align*}\tilde{p}\:.\tilde{p}=g^{\mu\nu}p_{\mu}p_{\nu}=p_{\mu}p^{\mu}=m^2u_{\mu}u^{\mu}=-m^2,\end{align*}
where 
\begin{align*}g^{\mu\nu}p_{\mu}p_{\nu}=g^{tt}{p_t}^2+2g^{t\phi}p_t p_\phi+g^{rr}p_r^2+g^{\phi\phi}p_\phi^2,\end{align*}
together with the results we obtained previously: $p_t=-E,p_\phi=L$, and the fact that $p_r=\frac{\partial \mathcal{L}}{\partial \dot{r}}=mg_{rr}\dot{r}$ , we can obtain a equation for $\dot{r}$.
Before doing so, recall the Kerr metric under contravariant basis $g^{\mu\nu}$ is just the inverse of the matrix $g_{\mu\nu}$, and this is 
\[
g^{\mu\nu} = 
\begin{pmatrix}
- \left(1 - \frac{2Mr}{\Sigma}\right)^{-1} & 0 & 0 & - \frac{2Mra}{\Sigma \left(1 - \frac{2Mr}{\Sigma}\right)} \\
0 & \frac{\Delta}{\Sigma} & 0 & 0 \\
0 & 0 & \frac{1}{\Sigma} & 0 \\
- \frac{2Mra}{\Sigma \left(1 - \frac{2Mr}{\Sigma}\right)} & 0 & 0 & \frac{\Sigma - 2Mr}{\Sigma \sin^2\theta \left(1 - \frac{2Mr}{\Sigma}\right)}
\end{pmatrix},
\]
c.f. \cite{1972ApJ...178..347B} [2.2]. Finally we can arrive at the radial time-like equatorial plane geodesic equation by substitute in $\dot{r}=\frac{g^{rr}}{m^2}(-m^2-g^{tt}E^2-2g^{t\phi}EL-g^{\phi\phi}L^2)$ and get:
\begin{equation}
\dot{r}=-\frac{\Delta}{r^2}+\frac{a^2+2a^2Mr^{-1}+r^2}{m^2r^2}E^2-\frac{4aM}{m^2r^3}EL-\frac{r-2M}{m^2r^3}L^2=:f(r).
\label{eq:r governing equation} 
\end{equation}
And further simplify equation\ref{eq:t geodesic equation} and \ref{eq:phi geodesic equation} using equatorial plane simplification $\theta=\pi/2$ , $p_{\theta}=\dot{\theta}=0$, we have
\begin{equation}
\dot{t}=-\frac{2aM}{m r\Delta}L+\frac{1}{m\Delta r^2}[ r^2(a^2+r^2)+2Ma^2r]E,
\label{eq:t governing equation} 
\end{equation}
\begin{equation}
\dot{\phi}=\frac{ r^2-2Mr}{m\Delta r^2}L+\frac{2aM}{m\Delta r}E.
\label{eq:phi governing equation} 
\end{equation}
Now we have obtained all four governing equations (with trivial polar coordinate equation $\theta=\frac{\pi}{2}$).

By rearranging and rewriting radial equation \ref{eq:r governing equation}, and define effective potential
\begin{align*}
V_{eff}(r):&=-(f(r)-(\frac{E^2}{m^2}-1))\\&=-\frac{2M}{r}+\frac{L^2-a^2E^2+m^2a^2}{m^2r^2}-\frac{2M(L-aE)^2}{m^2r^3},
\end{align*}
we can obtain the variant form of radial governing equation:
\begin{equation}
\dot{r}^2\:\:+\:\:V_{eff}(r)\:\:=\:\:\frac{E^2}{m^2}-1.
\label{eq: r governing equation with effective potential}
\end{equation}
\section{Circular orbits}
\label{Chapter: Circular orbit}
\begin{figure}
\centering
\includegraphics[width=0.75\linewidth]{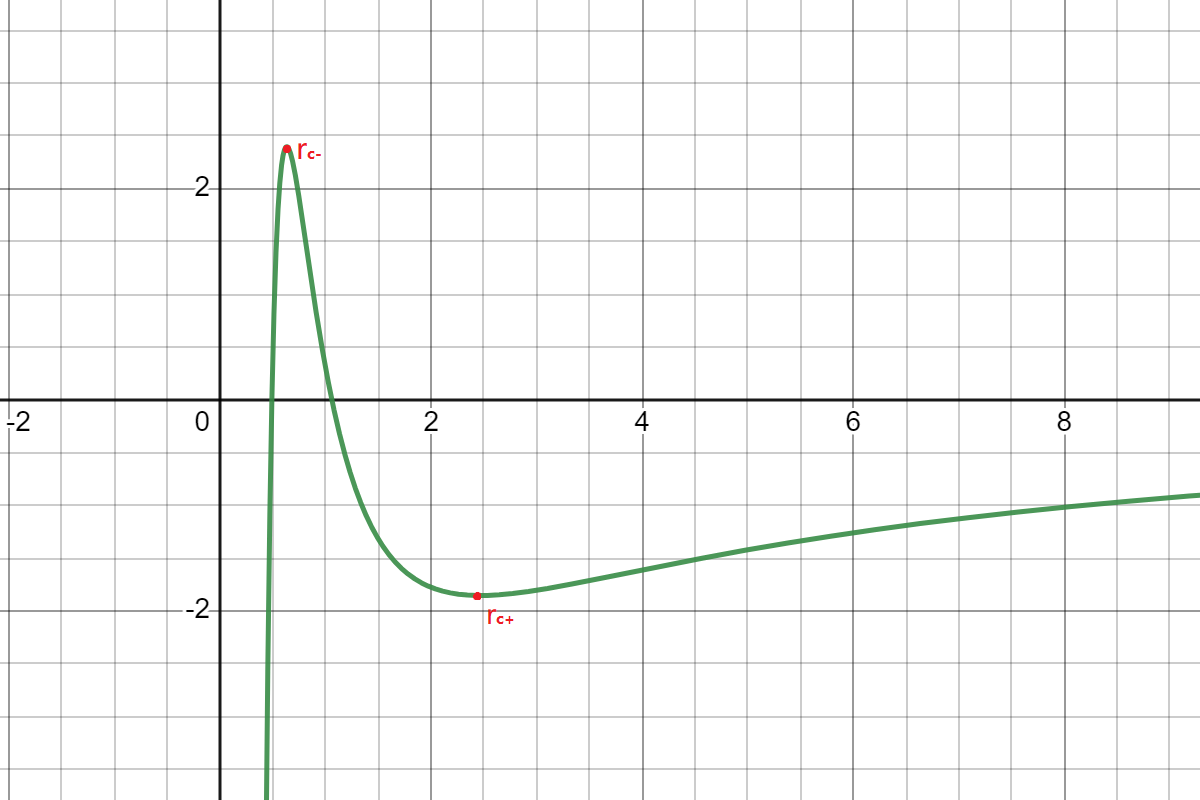}
\caption{\label{fig:Figure1}Plotting of effective potential \cite{DesmosGraph} }
\end{figure}
\subsection{Stable circular orbit(SCO)}
For a circular orbit, the first necessary condition is the zero radial movement, which means $\dot{r}=0$. Additionally, for a massive particle to maintain this orbit, it is also required that there is no radial acceleration, i.e.,  $\ddot{r}=0$.

We first look at condition $\ddot{r}=0$. By differentiating both sides of equation\ref{eq:r governing equation}, we can see $\ddot{r}=0$ implies $\frac{\text{d}f(r)}{\text{d}r}=0$. Or equivalently, $\frac{\text{d}V_{eff}(r)}{\text{d}r}=\frac{\text{d}f(r)}{\text{d}r}=0$ if we use \ref{eq: r governing equation with effective potential}. Solving these gives two solution of r:
\begin{align*}
    r_{c\pm}=\frac{L^2-a^2E^2+a^2m^2\pm\sqrt{(L^2-a^2E^2+a^2m^2)^2-12M^2m^2(L-aE)^2}}{2Mm^2}.
\end{align*}
As plotted in Figure \ref{fig:Figure1}, the reason of two roots is that we only required $\ddot{r}=0$ but not considered the sign of $\frac{\text{d}^2V_{eff}(r)}{\text{d}r^2}$.

By substitute $r_{c\pm}$ into $\frac{\text{d}^2V_{eff}(r)}{\text{d}r^2}$ , we will find:
\begin{align*}
\frac{\text{d}^2V_{eff}(r)}{\text{d}r^2}\lvert_{r=r_{c-}}\leqslant0 \:\:\forall E,L,\\\frac{\text{d}^2V_{eff}(r)}{\text{d}r^2}\lvert_{r=r_{c+}}\geqslant0 \:\:\forall E,L.
\end{align*}
Thus $r_{c+}$ is the stable orbit we are interested in, and $r_{c-}$ is the unstable orbit. And we will denote $r_{s}\equiv r_{stable}\equiv  r_{c+}$ from now on. 

So far we have taken $E,L$  as constants. However these can also be viewed as independent variable, if we want to answer what is the radius of stable circular orbit for any given $E,L$: 
\begin{align*}
r_s(E,L)=\frac{L^2-a^2E^2+a^2m^2+\sqrt{(L^2-a^2E^2+a^2m^2)^2-12M^2m^2(L-aE)^2}}{2Mm^2}=:G(E,L).
\end{align*}

Now we look at the $\dot{r}=0$ condition,  which implies $f(r)=0$. For a stable orbit, we have established that  $r=r_s$. Therefore we have $f(r_s)=f(r_s(E,L))=f_{r_s}(E,L)=0$ by combining with circular orbit condition. This condition indicates a relationship between $E$ (energy) and $L$ (angular momentum), suggesting that not all combinations of $E$ and $L$ result in a stable circular orbit. By solving the equation $f_{r_s}(E,L)=0$ we can obtain a relationship: either $E_s=E(L)$ as a function of $L$, or $L_s=L(E)$ as a function of $E$. 

In another word, a stable circular orbit must satisfy following conditions:
\begin{align*}
    \begin{cases}r_s=G(E,L) \\f(r_s(E,L))=0 &\end{cases}.
\end{align*}
Here, we have three variables — $E,L,r_s$ — and two equations. Solving these will yield a solution where, if one variable is known, the other two can be determined. In other words, knowing any one of $E,L$ or $r_s$  allows us to immediately find the remaining two quantities. 

For simplicity and further study, we will look at solution using $r_s$ as the free parameter and $E,L$ is determined by \cite{1972ApJ...178..347B} (2.12) (2.13):
\begin{equation}
    E_s = m\frac{r_s^{\frac{3}{2}} - 2Mr_s^{\frac{1}{2}} \pm aM^{\frac{1}{2}}}{r_s^{\frac{3}{4}}\left(r_s^{\frac{3}{2}} - 3Mr_s^{\frac{1}{2}} \pm 2aM^{\frac{1}{2}}\right)^{\frac{1}{2}}}\equiv E(r_s),
\label{eq:E stable circular orbit}
\end{equation}
\begin{equation}
    L_s = m\frac{\pm M^{\frac{1}{2}}(r_s^2 \mp 2aM^{\frac{1}{2}}r_s^{\frac{1}{2}} + a^2)}{r_s^{\frac{3}{4}}\left(r_s^{\frac{3}{2}} - 3Mr_s^{\frac{1}{2}} \pm 2aM^{\frac{1}{2}}\right)^{\frac{1}{2}}} \equiv L(r_s).
\label{eq:L stable circular orbit}
\end{equation}
\begin{align*}
    \text{Upper sign: Pro-grade Orbit(L $\geqslant$ 0); Lower sign: Retro-grade Orbit(L $\leqslant$ 0)} 
\end{align*}
These formulas tells us what must be the test particle's energy $E$ and angular momentum $L$, if this test particle travels on the SCO at $r=r_s$. 

And we substitute equations \ref{eq:E stable circular orbit} and \ref{eq:L stable circular orbit} in equations \ref{eq:t governing equation} and \ref{eq:phi governing equation} to get the time-dilation and azimuthal velocity at this SCO radius $r_s$:
\begin{align*}
    \dot{t}_s\equiv \dot{t}(r_s)= \dot{t}(E=E(r_s),L=L(r_s),r=r_s),\\
    \dot{\phi}_s\equiv \dot{\phi}(r_s)= \dot{\phi}(E=E(r_s),L=L(r_s),r=r_s).
\end{align*}
as well as the coordinate angular velocity at this SCO radius:
\begin{equation}
    \Omega_s\equiv \Omega(r_s)=\frac{\text{d}\phi}{\text{d}t}\lvert_{r=r_s}=\frac{\dot{\phi}_s}{\dot{t}_s}=\frac{\pm M^{\frac{1}{2}}}{r_s^{\frac{3}{2}}\pm aM^{\frac{1}{2}}}.
    \label{eq: coordinat angular velocity}
\end{equation}

\subsection{Innermost stable circular orbit(ISCO)}
So far we have solved all necessary equations with respect to $r_s$, that is once we know the basic information (m,M,a) and the radius of SCO $r_s$, we will have full information of the test particle dynamics(such as its time-dialation and angular velocity). 

During the process for solving for $r_s$, we ignored some boundary restrictions which we are going to pick up here. There appears to be a minimal possible stable circular orbit, which inside the limit there won't be any possible $r_s$, no matter how large the energy $E$ is. For example, any massive particle can not travel in a circular orbit inside or even near event horizon. Here we are going to derive the exact limit radius and this limit is called inner most stable orbit(ISCO).

We can analyse the ISCO by looking at solution for $r_s$. The solution of $r_s$ is only possible when $\sqrt{(L^2-a^2E^2+a^2m^2)^2-12M^2m^2(L-aE)^2}$ is valid, that is the ISCO is at where:
\begin{equation}
        (L^2-a^2E^2+a^2m^2)^2-12M^2m^2(L-aE)^2 = 0.
    \label{eq:isco reapeated root of r}
\end{equation}

Simplify and substitute \ref{eq:E stable circular orbit} and \ref{eq:L stable circular orbit} in we can obtain ISCO formula:
\begin{align*}
    L(r_{isco})^2-a^2E(r_{isco})^2+a^2m^2-\sqrt{12}Mm|L(r_{isco})-aE(r_{isco})|=0.
\end{align*}
Noticing here that $L(r)\propto m$  and $E(r)\propto m$ indicating each term in the equation contains a $m^2$ term, which can be all cancelled out and the formula is independent of $m$, indicating that the solution of $r_{isco}$ is actually independent of $m$. This is easy to understand intuitively as the ISCO should be a intrinsic property of the Black Hole itself, which is independent of the properties of test particles. Further more the difference in $m$ can  be `absorbed' into $E$ and $L$, as we can modify $E,L$ of the object coherently to offset the impact of changing in $m$.

And thus by solving this equation we can obtain some solution $r_{isco}$ , which is a constant that only depend on $a$ and $M$. The solution of $r_{isco}$ is \cite{1972ApJ...178..347B}:
\begin{equation}
    r_{isco}=M(3+Z_2\mp\sqrt{(3-Z_1)(3+Z_1+2Z_2)},
    \label{eq:r_isco solution}
\end{equation}
\begin{align*}
Z_1 &= 1+(1-\frac{a^2}{M^2})^{\frac{1}{3}}\left[(1+\frac{a}{M})^\frac{1}{3}+(1-\frac{a}{M})^\frac{1}{3}\right],\\
Z_2 &= \sqrt{3\frac{a^2}{M^2}+Z_1^2},
\end{align*}

\begin{align*}
    \text{Upper sign: Pro-grade Orbit(L $\geqslant$ 0); Lower sign: Retro-grade Orbit(L $\leqslant$ 0)} .
\end{align*}

It is note worthy that at ISCO, $(L^2-a^2E^2+a^2m^2)^2-12M^2m^2(L-aE)^2 = 0$ implies that $r_{c-}=r_{c+}$ , recall that $r_{c\pm}$ is the minimal stable/unstable circular orbit. So for massive test particles, $r_{isco}$ is not only the ISCO solution but also inner most unstable circular orbit or inner most circular orbit solution. And at $r=r_{isco}$, this is actually a saddle orbit. In fact when $E , L$ getting closer to $E_{isco}\equiv E(r_{isco}),L_{isco}\equiv L(r_{isco})$, the unstable and stable circular orbit given by the effective potential \ref{eq: r governing equation with effective potential} becomes closer and closer until they coincide. We will discuss these in detail at later chapters. 

\section{Spin-sign-converted Dimensionless (SSCD) formulas}
In this chapter, before delving deeper into the intricate aspects of Kerr trajectories, or even simply substitute $r_{isco}$ to get explicit $E_{isco}$, we seize an opportune moment to introduce the concept of `Spin-Sign-Converted Dimensionless Formulas.' These notations and the associated formulas are extensively employed in latest research, particularly in studies focused on equatorial plane dynamics. A pivotal reference in this domain is Ori-Thorne Transition procedure or Ori-Thorne-Kesden Transition procedure, which lays the foundational framework for understanding Kerr equatorial dynamics (see \cite{Finn_2000}\cite{Ori_2000}). Our previous discussions was fully based on standard Kerr notations and formulas as established by Bardeen (see \cite{1972ApJ...178..347B}). In this chapter, we aim to transition from standard formulations to Spin-Sign-Converted Dimensionless Formulas. This convention not only aligns with the evolving academic discourse but also reduce complexities when examining the nuanced dynamical formulas of Kerr spacetime. 

\subsection{Spin Sign Convention}
So far we have assumed $a=\frac{J}{M}$ to be positive when analysing Kerr equatorial trajectories, this led us to derive two distinct sets of equations based on the conditions $L \geqslant 0$ and $L \leqslant 0$, respectively. In this context, the condition $L \geqslant 0$ corresponds to `prograde' orbits, as the particle's motion is in the same direction as the black hole's spin. Conversely, $L \leqslant 0$ denotes `retrograde' orbits, where the particle moves in an orbit opposite to the black hole’s spin direction.

However, in this discussion, we will adopt a sign convention. We will set $L$ to be always positive and instead reflect the orbit's direction relative to the black hole's spin in the sign of $a$. This means that the sign of $a$ will now indicate whether the orbit is prograde or retrograde, and by doing this we can arrive at one set of equations, instead of two sets of equations representing prograde and retrograde individually.
\begin{align*}
    a^{(new)}&=sgn(aL)|a|=sgn(aL)a,\\
    L^{(new)}&=|L|=sgn(aL)L,
\end{align*}
where $sgn(aL)=sgn(L)=sgn(a^{new}L^{new})=sgn(a^{new})$.

Thus we substitute these into our previous equations, or by multiply both sides of equation by $sgn(aL)$, dropping “$^{(new)}$”. And thus obtains: 

\ref{eq:t governing equation} $\Longrightarrow$
\begin{equation}
\dot{t}=-\frac{2aM}{m r\Delta}L+\frac{1}{m\Delta r^2}[ r^2(a^2+r^2)+2Ma^2r]E;
\label{eq:t governing equation ssc} 
\end{equation}

\ref{eq:phi governing equation} $\Longrightarrow$
\begin{equation}
\dot{\phi}=sgn(a)\frac{ r^2-2Mr}{m\Delta r^2}L+sgn(a)\frac{2aM}{m\Delta r}E;
\label{eq:phi governing equation ssc} 
\end{equation}

\ref{eq:E stable circular orbit} $\Longrightarrow$
\begin{equation}
    E_s = m\frac{r_s^{\frac{3}{2}} - 2Mr_s^{\frac{1}{2}} + aM^{\frac{1}{2}}}{r_s^{\frac{3}{4}}\left(r_s^{\frac{3}{2}} - 3Mr_s^{\frac{1}{2}} + 2aM^{\frac{1}{2}}\right)^{\frac{1}{2}}}\equiv E(r_s);
\label{eq:E stable circular orbit ssc}
\end{equation}

\ref{eq:L stable circular orbit} $\Longrightarrow$
\begin{equation}
    L_s = m\frac{ M^{\frac{1}{2}}(r_s^2 - 2aM^{\frac{1}{2}}r_s^{\frac{1}{2}} + a^2)}{r_s^{\frac{3}{4}}\left(r_s^{\frac{3}{2}} - 3Mr_s^{\frac{1}{2}} + 2aM^{\frac{1}{2}}\right)^{\frac{1}{2}}} \equiv L(r_s);
\label{eq:L stable circular orbit ssc}
\end{equation}

\ref{eq: coordinat angular velocity} $\Longrightarrow$
\begin{align*}
    \Omega_s^{(new)}=\frac{\dot{\phi}_s^{(new)}}{\dot{t}_s^{(new)}}=\frac{sgn(aL)\dot{\phi}_s}{\dot{t}_s}=\frac{\pm sgn(aL) M^{\frac{1}{2}}}{r_s^{\frac{3}{2}}\pm sgn(aL)a^{(new)}M^{\frac{1}{2}}};
\end{align*}

$\Longrightarrow$
\begin{equation}
    \Omega_s\equiv \Omega(r_s)=\frac{ M^{\frac{1}{2}}}{r_s^{\frac{3}{2}}+ aM^{\frac{1}{2}}};
    \label{eq: coordinat angular velocity ssc}
\end{equation}

\ref{eq:r_isco solution} $\Longrightarrow$
\begin{equation}
    r_{isco}=M[{3+Z_2-sgn(a)\sqrt{(3-Z_1)(3+Z_1+2Z_2)}}],
    \label{eq:r_isco solution ssc}
\end{equation}
\begin{align*}
    with\:Z_1,Z_2\:unchanged;
\end{align*}
and equation \ref{eq: r governing equation with effective potential} remains invariant.

\subsection{Dimensionless}
Now we look at dimensionless notation, by defining dimensionless notations:
$\tilde{r}:=\frac{r}{M}$ ,$\tilde{t}:=\frac{t}{M}$, $\tilde{\phi}:=\frac{\phi}{M}$,$\tilde{a}:=\frac{a}{M}=\frac{J}{M^2}$, $\tilde{\tau}:=\frac{\tau}{M}$, $\eta:=\frac{m}{M}$, $\tilde{L}:=\frac{L}{mM}$ , $\tilde{E}=\frac{E}{m}$and $\tilde{\Omega}:=M\Omega$, $\tilde{\Delta}=\frac{\Delta}{M^2}$ \cite{Ori_2000} and substitute in the spin-sign convention notation and finally arrives at Spin-sign-converted Dimensionless formulas:
\begin{align*}
    \tilde{\Delta}=\tilde{r}^2-2\tilde{r}+\tilde{a}^2;
\end{align*}
\ref{eq:t governing equation ssc} $\Longrightarrow$time-coordinate SSCD governing equation:
\begin{equation}
\dot{\tilde{t}}=\frac{\text{d}\tilde{t}}{\text{d}\tilde{\tau}}=\frac{\text{d}t}{\text{d}\tau}=\dot{t}=-\frac{2\tilde{a}}{\tilde{r}\tilde{\Delta}}\tilde{L}+\frac{1}{\tilde{\Delta} \tilde{r}^2}[ \tilde{r}^2(\tilde{a}^2+\tilde{r}^2)+2\tilde{a}^2\tilde{r}]\tilde{E};
\label{eq:t governing equation SSCD} 
\end{equation}
similarly \ref{eq:phi governing equation ssc} $\Longrightarrow$azimuthal-coordinate SSCD governing equation:
\begin{equation}
\dot{\tilde{\phi}}=\dot{\phi}=sgn(\tilde{a})\frac{ \tilde{r}-2}{\tilde{\Delta}\tilde{r}}\tilde{L}+sgn(\tilde{a})\frac{2\tilde{a}}{\tilde{\Delta} \tilde{r}}\tilde{E};
\label{eq:phi governing equation SSCD} 
\end{equation}
\ref{eq:r governing equation} $\Longrightarrow$radial-coordinate SSCD governing equation:
\begin{equation}
    \dot{\tilde{r}}^2=\dot{r}^2=\tilde{E}^2-1+\frac{2}{\tilde{r}}-\frac{\tilde{L}^2-\tilde{a}^2\tilde{E}^2+\tilde{a}^2}{\tilde{r}^2}+\frac{2(\tilde{L}-\tilde{a}\tilde{E})^2}{\tilde{r}^3}=:f(\tilde{r});
    \label{eq: r governing equation SSCD}
\end{equation}
with effective potential now
\begin{align*}
V_{eff}(\tilde{r}):&=-(f(\tilde{r})-(\tilde{E}^2-1))\\&=-\frac{2}{\tilde{r}}+\frac{\tilde{L}^2-\tilde{a}^2\tilde{E}^2+\tilde{a}^2}{\tilde{r}^2}-\frac{2(\tilde{L}-\tilde{a}\tilde{E})^2}{\tilde{r}^3},
\end{align*}
and thus
\begin{equation}
\dot{\tilde{r}}^2\:\:+\:\:V_{eff}(\tilde{r})\:\:=\:\:\tilde{E}^2-1;
\label{eq: r governing equation with effective potential SSCD}
\end{equation}
\ref{eq:E stable circular orbit ssc} $\Longrightarrow$ SSCD Energy of corresponding dimensionless SCO orbit radius:
\begin{equation}
    \tilde{E}_s = \frac{1-\frac{2}{\tilde{r}_s}+\frac{\tilde{a}}{{\tilde{r}_s}^\frac{3}{2}}}{(1-\frac{3}{\tilde{r}_s}+\frac{2\tilde{a}}{{\tilde{r}_s}^\frac{3}{2}})^\frac{1}{2}}\equiv \tilde{E}(\tilde{r}_s);
\label{eq:E stable circular orbit SSCD}
\end{equation}
\ref{eq:L stable circular orbit ssc} $\Longrightarrow$SSCD Angular momentum of corresponding dimensionless SCO orbit radius:
\begin{equation}
    \tilde{L}_s = \frac{\tilde{r}^\frac{1}{2}_s-\frac{2\tilde{a}}{\tilde{r}_s}+\frac{\tilde{a}^2}{{\tilde{r}^\frac{3}{2}_s}}}{(1-\frac{3}{\tilde{r}_s}+\frac{2\tilde{a}}{{\tilde{r}^\frac{3}{2}}_s})^\frac{1}{2}}\equiv \tilde{L}(\tilde{r}_s);
\label{eq:L stable circular orbit SSCD}
\end{equation}
and \ref{eq: coordinat angular velocity ssc} $\Longrightarrow$ SSCD coordinate angular velocity of corresponding dimensionless SCO:
\begin{equation}
    \tilde{\Omega}_s\equiv \tilde{\Omega}(\tilde{r}_s)=M\Omega_s=\frac{1}{\tilde{r}_s^{\frac{3}{2}}+ \tilde{a}};
    \label{eq: coordinat angular velocity SSCD}
\end{equation}
\ref{eq:r_isco solution ssc} $\Longrightarrow$ SSCD ISCO radius solution:
\begin{equation}
    \tilde{r}_{isco}=3+Z_2-sgn(\tilde{a})\sqrt{(3-Z_1)(3+Z_1+2Z_2)},
    \label{eq:r_isco solution SSCD}
\end{equation}
with:
\begin{align*}
Z_1 &= 1+(1-\tilde{a}^2)^{\frac{1}{3}}[(1+\tilde{a})^\frac{1}{3}+(1-\tilde{a})^\frac{1}{3}],\\
Z_2 &= (3\tilde{a}^2+Z_1^2)^\frac{1}{2};
\end{align*}
and \ref{eq:E stable circular orbit SSCD} $\Longrightarrow$ SSCD Energy of particle at SSCD ISCO radius:
\begin{equation}
\tilde{E}_{isco}\equiv \tilde{E}(\tilde{r}_{isco}) = \frac{1-\frac{2}{\tilde{r}_{isco}}+\frac{\tilde{a}}{{\tilde{r}_{isco}}^\frac{3}{2}}}{(1-\frac{3}{\tilde{r}_{isco}}+\frac{2\tilde{a}}{{\tilde{r}_{isco}}^\frac{3}{2}})^\frac{1}{2}};
\label{eq:E isco SSCD}
\end{equation}

\ref{eq:L stable circular orbit SSCD} $\Longrightarrow$ SSCD Angular momentum of particle at dimensionless ISCO radius:
\begin{align*}
\tilde{L}_{isco} \equiv \tilde{L}(\tilde{r}_{isco})= \frac{\tilde{r}^\frac{1}{2}_{isco}-\frac{2\tilde{a}}{\tilde{r}_{isco}}+\frac{\tilde{a}^2}{{\tilde{r}^\frac{3}{2}_{isco}}}}{(1-\frac{3}{\tilde{r}_{isco}}+\frac{2\tilde{a}}{{\tilde{r}^\frac{3}{2}}_{isco}})^\frac{1}{2}};
\end{align*}
while we set $\tilde{a}=J/M^2$ , and use the fact that $J\leqslant M^2$ \footnote{The limitation $J\leqslant M^2$ for a black hole, where $J$ is the angular momentum and $M$ is the mass, ensures the existence of an event horizon in the Kerr black hole. This condition arises because, for a rotating (Kerr) black hole, the radius of the event horizon depends on both mass and angular momentum. Mathematically, the event horizon disappears if $J> M^2$, leaving a naked singularity, which is a theoretical anomaly where the central singularity of the black hole is exposed, violating the cosmic censorship hypothesis. Therefore, the constraint  $J\leqslant M^2$ is essential for the physical validity of a Kerr black hole, preserving the nature of the event horizon and aligning with our current understanding of the fundamental properties of black holes.} , then we have $-1\leqslant\tilde{a}\leqslant+1$ , and by using this domain and the result of ISCO radius we can further simplify $\tilde{L}_{isco}$ and obtains
\begin{equation}
\tilde{L}_{isco}=2\sqrt{3}-\frac{4\tilde{a}}{\sqrt{3\tilde{r}_{isco}}}
\label{eq:L isco SSCD},
\end{equation}
cf. \cite{Misner1973}.\\
\ref{eq:t governing equation SSCD} $\Longrightarrow$ SSCD local time-dilation at SSCD ISCO radius:
\begin{equation}
\dot{\tilde{t}}_{isco}=\dot{\tilde{t}}(\tilde{E}_{isco},\tilde{L}_{isco},\tilde{r}_{isco})=\frac{1+\frac{\tilde{a}}{\tilde{r}^{\frac{3}{2}}_{isco}}}{\sqrt{1-\frac{3}{\tilde{r}_{isco}}+\frac{2\tilde{a}}{{\tilde{r}^\frac{3}{2}}_{isco}}}},
\label{eq:t isco SSCD}
\end{equation}
cf. \cite{1973blho.conf..343N} (5.4.5.a).
\section{Adiabatic Inspiral and Plunge Regime}

Having derived all the necessary formulas, our next objective is to seek closed-form solutions using the equations. However, solving the full Kerr geodesics for intricate and dynamically evolving systems presents formidable challenges, due to inherent complexity arising from the non-linear nature of Kerr geodesics. Moreover, our primary interest lies in near-circular or slowly evolving particle orbits, as these scenarios are significantly more feasible to detect by LISA due to their prolonged and stable emission of gravitational waves. Therefore, our main focus will be on devising an approximation method that accurately captures the dynamics of such systems. The adiabatic inspiral approach aims to balance the need for computational efficiency with the requirement for sufficient accuracy to yield meaningful insights into the nature of orbit inspirals in the vicinity of rotating black holes. 

\subsection{ Gravitational Wave Emission }
It is crucial to recognise that in such binary systems, the energy and angular momentum of the test particle are no longer constants and there can not be any particle travels on circular orbit “forever”. This variation is primarily due to the emission of gravitational waves, which carry significant energy away from the system. 
The full detailed description of gravitational wave emission and calculation can be found in Chapter 36 \cite{Misner1973}. Here we will briefly summarise the derivation. 

To capture the gravitational wave energy we can borrow the idea used in electromagnetic theory. In Electromagnetic theory we use ``luminosity" $\mathbb{L}_{electirc-dipole}$ to represent the power output of electric radiation to lay down similar formula. While since there does not exist any gravitational dipole, and due to the gravitational wave has a quadruple nature, we can use based on power output predicted by electromagnetic theory to generate similar equation $\mathbb{L}_{GW}\equiv\mathbb{L}_{mass-quadraple}$ (36.1 in \cite{Misner1973})for gravitational wave. 

To capture gravitational wave energy, we can draw parallels with concepts from electromagnetic theory. In electromagnetic theory, the term "luminosity" $\mathbb{L}_{\text{electric-dipole}}$ represents the power output due to electric dipole radiation. This concept helps in formulating equations to describe the radiation's characteristics. However, for gravitational waves, the situation is different because gravitational interactions do not have a dipole moment due to the nature of gravity. Instead, gravitational waves are predominantly quadrupolar in nature, reflecting the fact that the lowest order non-vanishing multipole moment for gravity is the quadrupole. Therefore, to describe the power output of gravitational waves, we use an analogous concept adjusted for their quadrupole nature. This leads us to define a similar luminosity for gravitational waves, denoted as $\mathbb{L}_{\text{GW}}$, which is akin to the concept of mass quadrupole luminosity. This is formally introduced as $\mathbb{L}_{\text{GW}} \equiv \mathbb{L}_{\text{mass-quadrupole}}$, and its formulation can be found in Chapter 36.1, \cite{Misner1973}. It's important to note that while the analogy provides a useful framework, the specific mechanisms and formulations are distinct due to the fundamental differences between electromagnetic and gravitational phenomena.

For a binary system, which is our focus of interest, 
Kepler's laws dictate that two particles with mass $m_1$ and $m_2$ and with angular frequency $\Omega$ and separation $l$ follows the relation:
\begin{align*}
    \frac{(m_1+m_2)}{l^3}\equiv\frac{M}{l^3}=\Omega^2,
\end{align*}
where the kinetic energy here is
\begin{align*}
    KE=\frac{1}{2}m_1v_1^2+\frac{1}{2}m_2v_2^2=\frac{1}{2}\frac{m_1m_2}{m_1+m_2}l^2\Omega^2=\frac{1}{2}\frac{m_1m_2}{l}.
\end{align*}
The radiation power of the gravitational wave can be estimated roughly as the square of the circulating power:
\begin{align*}
\mathbb{L}_{GW}\sim \Omega \times (KE) = \frac{\mu^2 M^3}{4l^5}\mathbb{L}_0,
\end{align*}
where $\mu=m_1m_2/M$ is the reduced mass($\equiv$m for our previous discussion) , $M = m_1+m_2$ is the total mass($\equiv$M for our previous discussion). And $\mathbb{L}_0=\frac{c^5}{G}$ is the convention factor to convert expression into dimensionless formulas, which as our previous discussion this can be normalised to $c=G=\mathbb{L}_0=1$.

The exact solution from (36.1 \cite{Misner1973}) is
\begin{align*}
    \mathbb{L}_{GW}=\frac{32}{5}\frac{\mu^2 M^3}{l^5}f(\varepsilon)\mathbb{L}_0,
\end{align*}
given by (36.16 \cite{Misner1973}). We can see that it complies with what our derivation predicts. here $f(\epsilon)\equiv\dot{\epsilon}$ is the general relativistic correction to the Newtonian, again details are in (36.16 \cite{Misner1973}).

Therefore our energy loss rate is the radiation power:
\begin{align*}
    \text{d}E/\text{d}t= -\mathbb{L}_{GW} = -\frac{32}{5}\frac{\mu^2 M^3}{l^5}f(\varepsilon)\mathbb{L}_0,
\end{align*}
now replacing terms $l$ with $\Omega$ we have:
\begin{align*}
    \text{d}E/\text{d}t= -\frac{32}{5}\frac{\mu^2 M^3}{l^5}f(\varepsilon)\mathbb{L}_0\equiv-\frac{32}{5}\frac{\mu^2 M^3}{l^5}\dot{\varepsilon}\\
    =-\frac{32}{5}\frac{\mu^2 M^3}{(\frac{M}{\Omega^2})^\frac{5}{3}}\dot{\varepsilon}=-\frac{32}{5}\frac{\mu^2} {M^2}M^\frac{10}{3}\Omega^\frac{10}{3}\dot{\varepsilon}.
\end{align*}
And using the SSCD notation in our previous discussion, we then have our final energy lost rate formula:
\begin{equation}
    \text{d}\tilde{E}/\text{d}\tilde{t}=-\frac{32}{5}\eta\:\tilde{\Omega}^\frac{10}{3}\dot{\varepsilon}.
    \label{eq:energy lost rate}
\end{equation}

\begin{figure}
\centering
\includegraphics[width=0.85\linewidth]{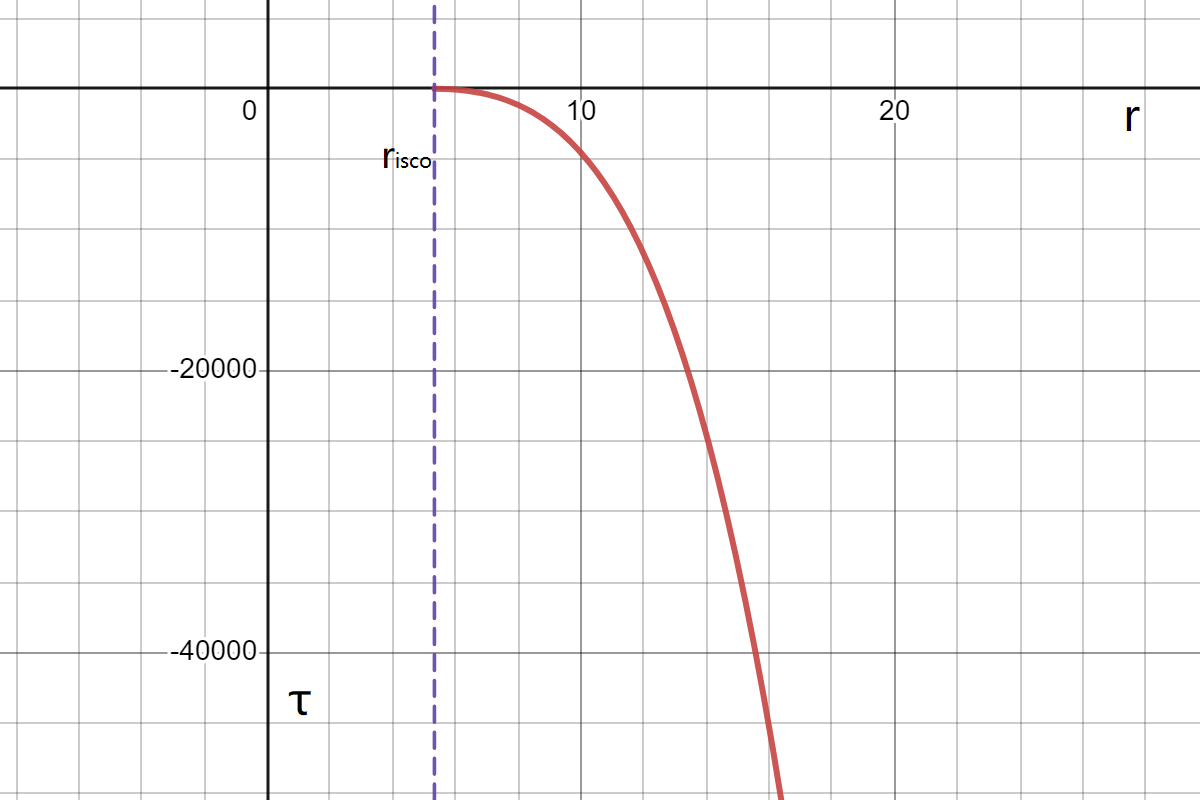}
\caption{\label{fig:Figure2}Plotting of Adiabatic Inspiral.  \cite{DesmosGraph} }
\includegraphics[width=0.65\linewidth]{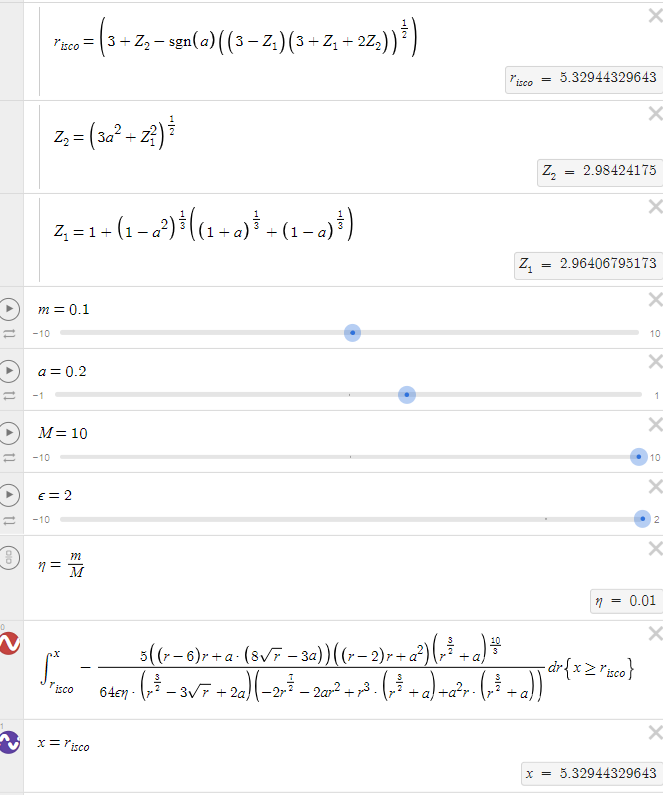}
\caption{\label{fig:Figure2.1}Parameters \cite{DesmosGraph} }
\end{figure}
\subsection{Timescale Separation }
To further discuss how we can construct an approximation, we first look at how does this gravitational-radiation energy lost affect object's trajectories. We will do this by comparing the timescale of circular orbit and gravitational-radiation.

The radiative timescale, $T_\text{rad}$, is the timescale over which energy is radiated away from the system, can be estimated by considering the amount of time it would take for a significant fraction of the total energy to be lost at the given rate:
\begin{align*}
    T_\text{rad}\approx \frac{E_s}{|\frac{dE}{dt}|}.
\end{align*}
Where using SSCD notation, this is 
\begin{align*}
    T_\text{rad}\approx \frac{\tilde{E}_s}{\frac{32}{5}\eta\:\tilde{\Omega}^\frac{10}{3}\dot{\varepsilon}}.
\end{align*}
The orbital timescale can then be estimated by comparing the total orbital energy to this power loss.
Where we have established angular frequency $\Omega$ for, so the orbital timescale can be estimated using orbital period:
\begin{align*}
    T_\text{orb}\approx \frac{2\pi}{\tilde{\Omega}}.
\end{align*}
Therefore, 
\begin{align*}
    \frac{T_\text{orb}}{T_\text{rad}}\approx \frac{2\pi}{\tilde{E}_s}{\frac{32}{5}\eta\:\tilde{\Omega}^\frac{7}{3}\dot{\varepsilon}}.
\end{align*}
Where $\tilde{E}_s \sim 1$ , $\tilde{\Omega}\ll1$ for $\tilde{r}_s \geqslant \tilde{r}_{isco}$. And $\eta=\frac{\mu}{M}\ll1$. Thus we can conclude that:
\begin{align*}
    \frac{T_{orb}}{T_{rad}}\ll1\:;\:T_{orb}\ll T_{rad}.
    \label{eq: Timescale}
\end{align*}

\subsection{Adiabatic Inspiral Regime}

From comparison\ref{eq: Timescale} we know that the orbital evolution driven by gravitational wave emission unfolds over a timescale that is significantly longer compared to the orbital period of a test particle. In other words, the particle completes many orbits before it loses a substantial amount of energy. This phenomenon indicates a slow inspiral, which is characteristic of adiabatic evolution. 

The adiabatic inspiral approximation is characterized by a slow and gradual change in the orbital parameters, or orbital evolution, as a consequence of gravitational wave emission. This process allows the system to adjust its energy and angular momentum progressively, ensuring that the orbital changes occur at a rate that is much slower than the rate of energy loss. As such, this approximation is crucial for understanding the dynamics of compact binary systems in the context of gravitational wave astronomy. 

In the previous calculations, we have seen that given any of three variables $E$, $L$ or $r_s$ allows for the determination of the remaining two. In the context of adiabatic inspiral, we select energy $E$ as the variable parameter, driven by the gravitational energy lost, and then examine the evolution of the orbital radius: 
\begin{align*}
    \frac{\text{d}\tilde{r}}{\text{d}\tilde{t}}=\frac{\text{d}\tilde{r}}{\text{d}\tilde{E}}\cdot\frac{\text{d}\tilde{E}}{\text{d}\tilde{t}}.
\end{align*}

Where equation \ref{eq:E stable circular orbit SSCD} $\Longrightarrow$
\begin{align*}
    \frac{\text{d}\tilde{r}}{\text{d}\tilde{E}}=1/\frac{\text{d}\tilde{E}}{\text{d}\tilde{r}}=\dfrac{2\left(-\frac{3}{ \tilde{r}}+\frac{2\tilde{a}}{ \tilde{r}^\frac{3}{2}}+1\right)^\frac{3}{2}}{\tilde{r}^{-2}-6 \tilde{r}^{-3}-3\tilde{a}^2 \tilde{r}^{-4}+8\tilde{a}\tilde{r}^{-\frac{7}{2}}}.
\end{align*}

And equations \ref{eq:energy lost rate} and \ref{eq:phi governing equation SSCD} $\Longrightarrow$
\begin{align*}
         \frac{\text{d}\tilde{r}}{\text{d}\tilde{t}}&=\frac{\text{d}\tilde{r}}{\text{d}\tilde{E}}\cdot-\frac{32}{5}\eta\:(\frac{1}{\tilde{r}^{\frac{3}{2}}+ \tilde{a}})^\frac{10}{3}\dot{\varepsilon}\\&=-\dfrac{64{\dot{\varepsilon}}{\eta}\cdot\left(-\frac{3}{\tilde{r}}+\frac{2\tilde{a}}{\tilde{r}^\frac{3}{2}}+1\right)^\frac{3}{2}}{5\left(\frac{1}{\tilde{r}^2}-\frac{6}{\tilde{r}^3}+\frac{8\tilde{a}}{\tilde{r}^\frac{7}{2}}-\frac{3\tilde{a}^2}{\tilde{r}^4}\right)\left(\tilde{r}^\frac{3}{2}+\tilde{a}\right)^\frac{10}{3}}.
\end{align*}
Then by switching from coordinate time to proper time using equation \ref{eq:t governing equation SSCD}, we can conclude that: 
\begin{gather*}
    \frac{\text{d}\tilde{r}}{\text{d}\tilde{\tau}}=\frac{\text{d}\tilde{r}}{\text{d}\tilde{t}}\cdot\frac{\text{d}\tilde{t}}{\text{d}\tilde{\tau}}\\
    =-\dfrac{64{\dot{\varepsilon}}{\eta}\cdot\left(\tilde{r}^\frac{3}{2}-3\sqrt{\tilde{r}}+2\tilde{a}\right)\left(-2\tilde{r}^\frac{7}{2}-2\tilde{a}\tilde{r}^2+\tilde{r}^3\cdot\left(\tilde{r}^\frac{3}{2}+\tilde{a}\right)+\tilde{a}^2\tilde{r}\cdot\left(\tilde{r}^\frac{3}{2}+\tilde{a}\right)\right)}{5\left(\left(\tilde{r}-6\right)\tilde{r}+\tilde{a}\cdot\left(8\sqrt{\tilde{r}}-3\tilde{a}\right)\right)\left(\left(\tilde{r}-2\right)\tilde{r}+\tilde{a}^2\right)\left(\tilde{r}^\frac{3}{2}+\tilde{a}\right)^\frac{10}{3}}.
\end{gather*}
We can see that we have established an ODE that $\frac{\text{d}\tilde{r}}{\text{d}\tilde{\tau}}=$ some function of $\tilde{r}$, we can therefore obtain the adiabatic inspiral relation:
\begin{align*}
    \tilde{\tau}=\int_{\tilde{r}_0}^{\tilde{r}}\frac{\text{d}\tilde{\tau}}{\text{d}\tilde{r}'}\text{d}\tilde{r}',
\end{align*}
once we pick the proper $\tilde{r}_0$. Where we have previously seen that the circular orbit only exists for $\tilde{r}\geqslant\tilde{r}_{isco}$, thus our adiabatic inspiral approximation will only work outside the ISCO. And in order to further set up discussion of plunge and transition regime, it is convenient to set $\tilde{\tau}=0$ at $\tilde{r}=\tilde{r}_{isco}$ . Thus we will set $\tilde{r}_0:=\tilde{r}_{isco}$ here, and finally obtain the adiabatic governing equation:
\begin{equation}
    \tilde{\tau}(\tilde{r})=\int_{\tilde{r}_{isco}}^{\tilde{r}}-\dfrac{5\left(\left(\tilde{r}'-6\right)\tilde{r}'+\tilde{a}\cdot\left(8\sqrt{\tilde{r}'}-3\tilde{a}\right)\right)\left(\left(\tilde{r}'-2\right)\tilde{r}'+\tilde{a}^2\right)\left(\tilde{r}'^{\frac{3}{2}}+\tilde{a}\right)^{\frac{10}{3}}}{64{\dot{\varepsilon}}{\eta}\cdot\left(\tilde{r}'^{\frac{3}{2}}-3\sqrt{\tilde{r}'}+2\tilde{a}\right)\left(-2\tilde{r}'^{\frac{7}{2}}-2\tilde{a}\tilde{r}'^2+\tilde{r}'^3\cdot\left(\tilde{r}'^{\frac{3}{2}}+\tilde{a}\right)+\tilde{a}^2\tilde{r}'\cdot\left(\tilde{r}'^{\frac{3}{2}}+\tilde{a}\right)\right)} \text{d}\tilde{r}'
    \label{eq: Adiabatic}
\end{equation}
with domain
\begin{align*}
    \tilde{r}\geqslant\tilde{r}_{isco}    .
\end{align*}
Figure \ref{fig:Figure2} shows the plot of adiabatic governing equation \ref{eq: Adiabatic}, which one can find either from plot or deriving from equation \ref{eq: Adiabatic} that it fails around $\tilde{r}_{isco}$, due to $\frac{\text{d}\tilde{r}}{\text{d}\tilde{\tau}}\longrightarrow-\infty$ . We will discuss about this later.

Solving the inverse of \ref{eq: Adiabatic} and further substituting into the previous SSCD equations, we can then get the full dynamical solution of adiabatic inspiral with respect to proper time. 

\subsection{Plunge Regime}
\begin{figure}
\centering
\includegraphics[width=0.65\linewidth]{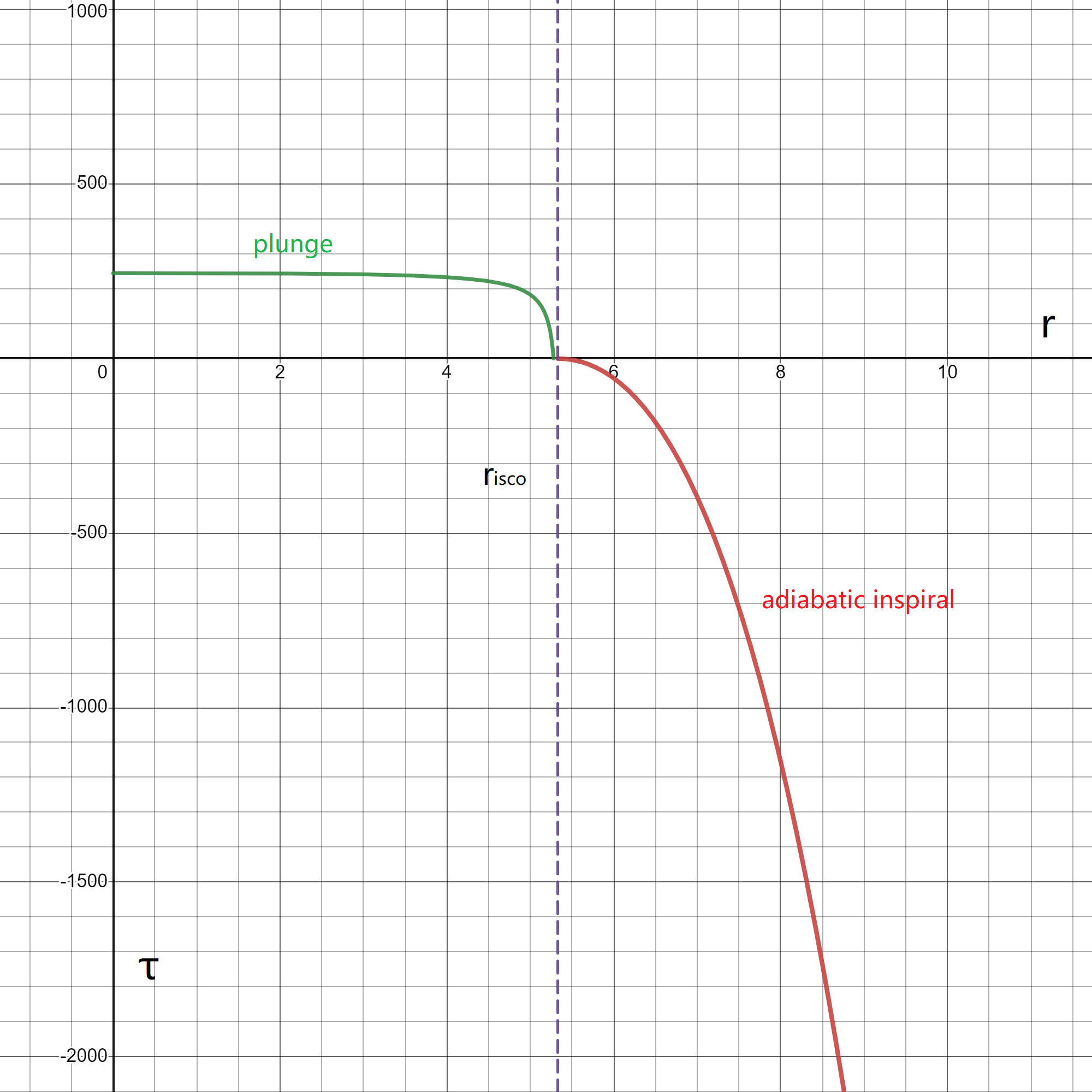}
\caption{\label{fig:Figure3}Plotting of Plunge together with Adiabatic Inspiral \cite{DesmosGraph} }
\includegraphics[width=0.65\linewidth]{AdiabaticInspiralParam.PNG}
\caption{\label{fig:Figure3.1}Parameters appending to Figure \ref{fig:Figure2.1} \cite{DesmosGraph} }
\end{figure}

In our previous discussion we have established a qualitative approximation for the trajectory of a massive particle following a near-circular orbit at the equatorial plane of a Kerr black hole, located outside the innermost stable circular orbit. We now turn our attention to the dynamics within the innermost stable orbit, an area known as the plunge regime.

For the analysis of this plunge regime, we adopt the concept of `action-free geodesic motion' as described in \cite{Misner1973}. This approach treats both energy and angular momentum as constants. Employing a process similar to the one used in our discussion of the adiabatic inspiral, one can demonstrate that within the plunge regime, the rate of energy loss has an almost negligible timescale compared to the timescale of the plunging motion itself, meaning we can again neglect any gravitational energy lost and treat E and L as constants. It is important to note that while the particle continues to lose energy during the plunge, this loss is relatively insignificant when compared to the rapidity of its descent into the black hole.

Here we use the constants $\tilde{E}:=\tilde{E}_{isco}$, $\tilde{L}:=\tilde{L}_{isco}$ . That is we assume that the energy and angular momentum remain same as the ISCO energy and angular momentum after the particle crosses the ISCO and enters the plunge regime.

Thus our governing equation would be \ref{eq: r governing equation SSCD} $\Longrightarrow$
\begin{align*}
    \dot{\tilde{r}}^2=\tilde{E}_{isco}^2-1+\frac{2}{\tilde{r}}-\frac{\tilde{L}_{isco}^2-\tilde{a}^2\tilde{E}_{isco}^2+\tilde{a}^2}{\tilde{r}^2}+\frac{2(\tilde{L}_{isco}-\tilde{a}\tilde{E}_{isco})^2}{\tilde{r}^3}=:f'(\tilde{r}),
\end{align*}
 and we can obtain the plunge relation by integrating both sides w.r.t. $\tilde{\tau}$:
\begin{align*}
    \tilde{\tau}(\tilde{r})=\int_{\tilde{r}}^{\tilde{r}_0}\frac{1}{\sqrt{f'(\tilde{r}')}}\text{d}\tilde{r}',
\end{align*}
with some suitable integrating constants $\tilde{r}_0$ similarly as when we solve adiabatic inspiral. But here, despite this relation should works for all values of radius $\tilde{r}\leqslant\tilde{r}_{isco}$ , it is not appropriate to take $\tilde{r}$ as $\tilde{r}_{isco}$ here. The reason is that in our assumption we used $\tilde{E}_{isco}$ and $\tilde{L}_{isco}$ as Energy and Angular momentum constants, however $\tilde{r}=\tilde{r}_{isco}$ is in fact the solution of stable circular orbit of ISCO given $\tilde{E}_{isco}$ and $\tilde{L}_{isco}$. Thus if $\tilde{r}$ is taken as $\tilde{r}_{isco}$, then we will get a trivial $\tilde{\tau}(\tilde{r})$ as the radial motion equation is $\tilde{r}=constant$ . We shall take a even more closer look at it:\\
from SCO governing equation $\Longrightarrow$ \ref{eq:E stable circular orbit SSCD} and \ref{eq:L stable circular orbit SSCD} solves \ref{eq: r governing equation SSCD} i.e.,
\begin{align*}
    f(\tilde{E}(\tilde{r}),\tilde{L}(\tilde{r}),\tilde{r})=0.
\end{align*}
Thus
\begin{align*}
    f'(\tilde{r}_{isco})=f(\tilde{E}_{isco},\tilde{L}_{isco},\tilde{r}_{isco})=f(\tilde{E}(\tilde{r}_{isco}),\tilde{L}(\tilde{r}_{isco}),\tilde{r}_{isco})=0.
\end{align*}
And therefore 
 \begin{align}
     \frac{1}{\sqrt{f'(r)}}\text{ is singular at } r= \tilde{r}_{isco},
 \end{align}
and 
 \begin{align*}
    \int_{c < \tilde{r}_{isco}}^{\tilde{r}_{isco}-\delta}\frac{1}{\sqrt{f'(\tilde{r}')}}\text{d}\tilde{r}'\longrightarrow \infty \text{ as } \delta\longrightarrow0.
\end{align*}
Thus we cannot even choose a $\tilde{r}_0 \sim \tilde{r}_{isco}$ . Intuitively, this is because given the energy and angular momentum settings as above, if the test particle is placed at ISCO, then it will stay there forever. Or even place at near-ISCO radius it will follows a almost-circular orbit for significant amount of time before enters proper plunge regime, i.e. $\Delta \tilde{r}<<\tilde{r}$. We will solve this by assuming that this particle has already crossed the ISCO, by giving it a `small inward push' at isco by setting $\tilde{r}_0:=\tilde{r}_{isco}-\delta$ , and thus obtains the final plunge relation:
\begin{equation}
    \tilde{\tau}(\tilde{r})=\int_{\tilde{r}}^{\tilde{r}_{isco}-\delta}\frac{1}{\sqrt{\tilde{E}_{isco}^2-1+\frac{2}{\tilde{r}'}-\frac{\tilde{L}_{isco}^2-\tilde{a}^2\tilde{E}_{isco}^2+\tilde{a}^2}{\tilde{r}'^2}+\frac{2(\tilde{L}_{isco}-\tilde{a}\tilde{E}_{isco})^2}{\tilde{r}'^3}}}\text{d}\tilde{r}'.
    \label{eq: plunge governing equation}
\end{equation}

\section{Transition Regime}
\subsection{Ori-Thorne Transition Procedure}
\begin{figure}
\centering
\includegraphics[width=0.65\linewidth]{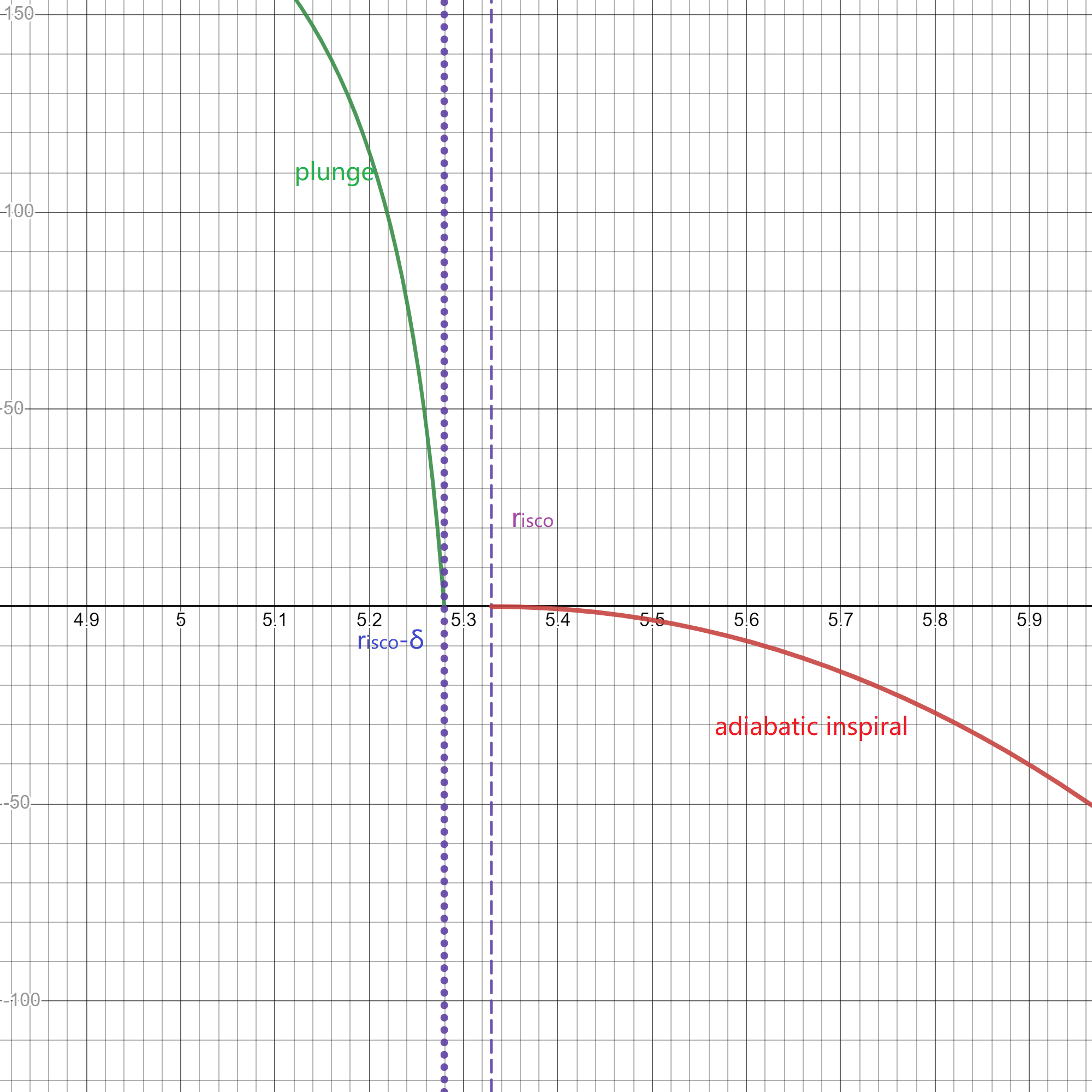}
\caption{\label{fig:Figure4}Zoom in of plunge and adiabatic plotting \cite{DesmosGraph} }
\end{figure}
In our analysis thus far, we have derived satisfactory approximate solutions for a massive test particle traveling in a near-circular orbit around a Kerr black hole initially. These approximations are valid under specific conditions: for orbits much larger than the innermost stable circular orbit (ISCO), i.e., $\tilde{r}>>\tilde{r}_{isco}$, and for orbits significantly smaller than the ISCO, i.e., $\tilde{r}<<\tilde{r}_{isco}$. However, they fail to accurately describe the particle's motion when the orbital radius is comparable to the ISCO radius, $\tilde{r}\sim\tilde{r}_{isco}$.

The plunge approximation becomes inadequate primarily because it assumes the particle’s energy and angular momentum are constant and equal to those at the ISCO. This implies that the plunge regime must originate from a position inside the ISCO, which has to be strictly smaller than ISCO. Similarly, the adiabatic approximation breaks down as the particle approaches the ISCO. Where the particle has acquired a significant radial velocity from inspiral, violating the assumption of a slow and gradual change in the orbital radius, i.e., when $\Delta \tilde{r}\sim \tilde{r}$, and the timescale of the orbit becomes comparable to the timescale of gravitational radiation-induced energy loss.

These limitations near the ISCO suggest the necessity of a `transition' period. This transition would smoothly connect the adiabatic inspiral phase with the plunge phase, providing a more comprehensive understanding of the particle's trajectory in this critical region. 

There are varies way to make some assumptions to simplify what happens near ISCO, and then give an solution for transition regime. Here we will fist look at Ori-Thorne transition procedure \cite{Ori_2000}. The Ori-Thorne analysis of the transition regime is confined to quasi-circular, equatorial orbits of massive particle, and we will see how these assumptions were utilised and if they have any restriction.

\subsubsection{Ori-Thorne Transition Approximations}
\label{Chapter: OT Approximation}
To effectively analyse the transition regime between adiabatic inspiral and plunge around a Kerr black hole, we begin by synthesizing appropriate assumptions. This involves adopting all reasonable assumptions and refining those that do not hold across the adiabatic and plunge phases. 

During the adiabatic inspiral, we assumed that energy is gradually dissipated over time due to the emission of gravitational waves, and the orbiting object maintains a nearly perfect circular trajectory, as the timescale of energy loss is significantly longer than the orbital period. Conversely, in the plunge phase, it is assumed that the loss of energy and angular momentum is negligible—effectively treated as constant—owing to a timescale of loss that greatly exceeds the duration of the plunge, and the object is no longer restricted on SCO and allow deviation from strict circular orbits. 

For the Ori-Thorne transition procedure, we propose that energy continues to be lost over time, albeit at a rate influenced by the proximity to the innermost stable circular orbit, and while the object's motion becomes increasingly unconstrained, it still follows trajectories that approximate circular orbits during transition regime, to reflect a complex interplay between gravitational radiation-driven decay and the intrinsic dynamics of near-ISCO orbits. 

\subsubsection{Energy-Angular Momentum relation of near-circular trajectory}

For an massive particle travelling on a near-circular trajectory, the ratio of energy radiation to angular momentum radiation equals to the body's orbital angular velocity, as given by \cite{1971reas.book.....Z} (10.7.22):
\begin{align*}
    \delta E= \Omega \delta L,
\end{align*}
i.e.
\begin{align*}
    \frac{\text{d}E}{\text{d}\tau}=\Omega\frac{\text{d}L}{\text{d}\tau},
\end{align*}
and switching to SSCD notation simply we can obtain the energy-angular momentum relation
\begin{equation}
       \frac{\text{d}E}{\text{d}\tau}\frac{m}{M}=\Omega\frac{\text{d}L}{\text{d}\tau}\frac{mM}{M^2}\Longrightarrow\frac{\text{d}\tilde{E}}{\text{d}\tilde{\tau}}=\tilde{\Omega}\frac{\text{d}\tilde{L}}{\text{d}\tilde{\tau}}.
       \label{eq:OT Transition regime E-L relation}
\end{equation}

It is important to note that the aforementioned relation, which equates the ratio of energy loss to angular momentum loss with the orbital angular velocity, assumes the object is in a near-circular orbit. This assumption may not hold in the latter half of the transition regime, where the object's motion transitions towards an almost-plunge phase, deviating significantly from circularity. However during the plunge phase of the transition regime, the impact of energy loss on the trajectory becomes increasingly negligible, minimizing its influence on the dynamics of the plunge itself. Therefore, while the initial assumption of near-circular orbits may become less accurate as the object approaches the plunge phase, the diminishing effect of energy loss on the overall dynamics allows us to extend the initial framework through the entire transition regime.

\subsubsection{Energy and Angular Momentum Loss Near ISCO}

During our discussion in previous chapter, we have seen the explicit formula for gravitational-wave-induced energy loss equation of near-circular orbit particles. This equation will still be used here but we will add some assumptions and simplifications for further need of transition regime.

We first convert the energy lost in terms of coordinate time to energy lost in terms of local proper time:
 \ref{eq:t governing equation SSCD}, \ref{eq:energy lost rate} $\Longrightarrow$
\begin{align*}
\frac{\text{d}\tilde{E}}{\text{d}\tilde{\tau}}&=\frac{\text{d}\tilde{E}}{\text{d}\tilde{t}}\cdot\frac{\text{d}\tilde{t}}{\text{d}\tilde{\tau}}\\
&=-\frac{32}{5}\eta\:\tilde{\Omega}^\frac{10}{3}\dot{\varepsilon}\dot{\tilde{t}}.
\end{align*}
In the Ori-Thorne transition procedure, we assume that the particle is travelling on quasi-circular orbits near ISCO with $\eta<<1$, which keeps its gravitational radiation weak. Thus its angular velocity and time-dilation are similar to ISCO constants. Thus we take:
\begin{align*}
    \tilde{\Omega}\simeq\tilde{\Omega}_{isco},\dot{\tilde{t}}\simeq\dot{\tilde{t}}_{isco},
\end{align*}
and then we will have:
\begin{align*}
        \frac{\text{d}\tilde{E}}{\text{d}\tilde{\tau}}=\frac{\text{d}\tilde{E}}{\text{d}\tilde{\tau}}\lvert_{r=r_{isco}}=constant=-\frac{32}{5}\eta\:\tilde{\Omega}_{isco}^\frac{10}{3}\dot{\varepsilon}\dot{\tilde{t}}_{isco}.
\end{align*}
Substitute in equation \ref{eq:t isco SSCD} that is:
\begin{equation}
    \frac{\text{d}\tilde{E}}{\text{d}\tilde{\tau}}=-\kappa\eta\tilde{\Omega}_{isco},\: \kappa:=\frac{32}{5}\:\tilde{\Omega}_{isco}^\frac{7}{3}\dot{\varepsilon}\frac{1+\frac{\tilde{a}}{\tilde{r}^{\frac{3}{2}}_{isco}}}{\sqrt{1-\frac{3}{\tilde{r}_{isco}}+\frac{2\tilde{a}}{{\tilde{r}^\frac{3}{2}}_{isco}}}}.
    \label{eq:OT Transition Kappa}
\end{equation}
So energy and angular momentum change is therefore:
\begin{align*}
    \tilde{E}(\tilde{\tau})&=-\kappa\eta\tilde{\Omega}_{isco}\tilde{\tau}+\tilde{E}_{0},\\
    \tilde{L}(\tilde{\tau})&=-\kappa\eta\tilde{\tau}+\tilde{L}_{0},
\end{align*}
with integration constant $\tilde{E}_0,\tilde{L}_0$. For consistency with our previous discussion, we would like to take $\tilde{\tau}=0$ at $\tilde{r}=\tilde{r}_{isco}$, and then we arrives at:
\begin{equation}
            \tilde{E}(\tilde{\tau})=-\kappa\eta\tilde{\Omega}_{isco}\tilde{\tau}+\tilde{E}_{isco},
            \label{eq:OT transition E lost rate}
\end{equation}
\begin{equation}
    \tilde{L}(\tilde{\tau})=-\kappa\eta\tilde{\tau}+\tilde{L}_{isco}.
    \label{eq:OT transition L lost rate}
\end{equation}
\subsubsection{Governing Equation}
\label{eq:Veff expansion}
For the fundamental radial governing equation, we start from equation \ref{eq: r governing equation SSCD}, with $\tilde{E}$ and $\tilde{L}$ treated as variables: 
\begin{align*}
    \dot{\tilde{r}}^2=f(\tilde{E},\tilde{L},\tilde{r})=\tilde{E}^2-1+\frac{2}{\tilde{r}}-\frac{\tilde{L}^2-\tilde{a}^2\tilde{E}^2+\tilde{a}^2}{\tilde{r}^2}+\frac{2(\tilde{L}-\tilde{a}\tilde{E})^2}{\tilde{r}^3}.
\end{align*}

Noticing here that if one can obtain some reasonable formula for $\tilde{E}$ and $\tilde{L}$ as either constant or function of $\tilde{r}$ and $\tilde{\tau}$ , then we will arrive at an ODE that is ready to be solved. Using the Ori-Thorne approximation of near ISCO energy and angular momentum loss, i.e. substitute equations \ref{eq:OT transition L lost rate},\ref{eq:OT Transition regime E-L relation} in, we therefore have:
\begin{align*}
    \dot{\tilde{r}}^2=f(\tilde{E}(\tilde{\tau}),\tilde{L}(\tilde{\tau}),\tilde{r})=f(\tilde{\tau},\tilde{r}).
\end{align*}

This ODE is complicated and we start by using techniques to simplify this, first differentiating with respect to $\tilde{\tau}$ in both sides:
\begin{align*}
    2\dot{\tilde{r}}\ddot{\tilde{r}}=\frac{\partial f}{\partial \tilde{\tau}}+\dot{\tilde{r}}\frac{\partial f}{\partial \tilde{r}}.
\end{align*}

The term $\frac{\partial f}{\partial \tilde{\tau}}$was neglected by Ori-Thorne here. We will call this term \textbf{time-dependent radial background force\label{name: timedependent radial background force}} , as it corresponds to the change of potential w.r.t. time. Ori-Thorne mentioned that the is slowly evolving \cite{Ori_2000}, thus it should be negligible comparing to the force from static background potential gradient corresponding to the $\frac{\partial f}{\partial \tilde{r}}$ term.

Thus we choose to neglect this time-dependent background force and thus obtain:
\begin{align*}
    \ddot{\tilde{r}}=\frac{1}{2}\frac{\partial f(\tilde{\tau},\tilde{r})}{\partial \tilde{r}}=-\frac{1}{2}\frac{\partial V_{eff}(\tilde{\tau},\tilde{r})}{\partial \tilde{r}}.
\end{align*}
Using the $V_{eff}$ from equation \ref{eq: r governing equation with effective potential SSCD}. 

Taylor expanding around ISCO, i.e. around $(\tilde{\tau}=0,\tilde{r}=\tilde{r}_{isco})$ up to the cubic order gives
\begin{align*}
&V_{\text{eff}}(\tilde{\tau}, \tilde{r}) = V_{\text{eff}}(0, \tilde{r}_{\text{isco}}) \\&+ \frac{\partial V_{\text{eff}}}{\partial \tilde{\tau}}(0, \tilde{r}_{\text{isco}})\cdot\tilde{\tau} + \frac{\partial V_{\text{eff}}}{\partial \tilde{r}}(0, \tilde{r}_{\text{isco}})\cdot(\tilde{r} - \tilde{r}_{\text{isco}}) \\&+ \frac{1}{2!}\left(\frac{\partial^2 V_{\text{eff}}}{\partial \tilde{\tau}^2}(0, \tilde{r}_{\text{isco}})\cdot\tilde{\tau}^2 + 2\frac{\partial^2 V_{\text{eff}}}{\partial \tilde{\tau} \partial \tilde{r}}(0, \tilde{r}_{\text{isco}})\cdot\tilde{\tau}(\tilde{r} - \tilde{r}_{\text{isco}}) + \frac{\partial^2 V_{\text{eff}}}{\partial \tilde{r}^2}(0, \tilde{r}_{\text{isco}})\cdot(\tilde{r} - \tilde{r}_{\text{isco}})^2\right)
\\&+ \frac{1}{3!}[\frac{\partial^3 V_{\text{eff}}}{\partial \tilde{\tau}^3}(0, \tilde{r}_{\text{isco}})\cdot\tilde{\tau}^3 + 3\frac{\partial^3 V_{\text{eff}}}{\partial \tilde{\tau}^2 \partial \tilde{r}}(0, \tilde{r}_{\text{isco}})\cdot\tilde{\tau}^2(\tilde{r} - \tilde{r}_{\text{isco}}) \\&+ 3\frac{\partial^3 V_{\text{eff}}}{\partial \tilde{\tau} \partial \tilde{r}^2}(0, \tilde{r}_{\text{isco}})\cdot\tilde{\tau}(\tilde{r} - \tilde{r}_{\text{isco}})^2 + \frac{\partial^3 V_{\text{eff}}}{\partial \tilde{r}^3}(0, \tilde{r}_{\text{isco}})\cdot(\tilde{r} - \tilde{r}_{\text{isco}})^3] \\&+ \cdots
\end{align*}
Where we have know some facts from earlier derivation in Chapter \ref{Chapter: Circular orbit}, that:\\
$\tilde{r}_{isco}$ is a solution of circular orbit $\Longrightarrow$
\begin{align*}
    V_{\text{eff}}(0, \tilde{r}_{\text{isco}}) = 1-\tilde{E}^2_{isco},\frac{\partial V_{\text{eff}}}{\partial \tilde{r}}(0, \tilde{r}_{\text{isco}})=0.
\end{align*}
$\tilde{r}_{isco}$ is a saddle point, or equivalently the repeated root of equation\ref{eq:isco reapeated root of r}  $\Longrightarrow$
\begin{align*}
    \frac{\partial^2 V_{\text{eff}}}{\partial \tilde{r}^2}(0, \tilde{r}_{\text{isco}})=0.
\end{align*}
The linearity of $E(\tilde{\tau})$, $L(\tilde{\tau})$ and thus quadraticity of $f$, $V_{eff}$ w.r.t $\tilde{\tau}$ $\Longrightarrow$
\begin{align*}
     \frac{\partial^2 V_{\text{eff}}}{\partial \tilde{\tau}^2} = constant.
\end{align*}
Therefore,
\begin{align*}
    \frac{\partial^3 V_{\text{eff}}}{\partial \tilde{\tau}^3}(0, \tilde{r}_{\text{isco}})=\frac{\partial^3 V_{\text{eff}}}{\partial \tilde{\tau}^2 \partial \tilde{r}}(0, \tilde{r}_{\text{isco}})=0.
\end{align*}
Further by the smoothness of $f$ ( $r\neq 0$) and Schwartz's theorem:
\begin{align*}
    \frac{\partial^3 V_{\text{eff}}}{\partial \tilde{r} \partial \tilde{\tau}^2 }(0, \tilde{r}_{\text{isco}})=0,
\end{align*}
Thus now we are left with:
\begin{align*}
    V_{\text{eff}}(\tilde{\tau}, \tilde{r}) &= 1-\tilde{E}^2_{isco}+ \frac{\partial V_{\text{eff}}}{\partial \tilde{\tau}}(0, \tilde{r}_{\text{isco}})\cdot\tilde{\tau}  \\&+\frac{1}{2!}\left(\frac{\partial^2 V_{\text{eff}}}{\partial \tilde{\tau}^2}(0, \tilde{r}_{\text{isco}})\cdot\tilde{\tau}^2 + 2\frac{\partial^2 V_{\text{eff}}}{\partial \tilde{\tau} \partial \tilde{r}}(0, \tilde{r}_{\text{isco}})\cdot\tilde{\tau}(\tilde{r} - \tilde{r}_{\text{isco}}) \right) \\&+\frac{1}{3!}( \frac{\partial^3 V_{\text{eff}}}{\partial \tilde{r}^3}(0, \tilde{r}_{\text{isco}})\cdot(\tilde{r} - \tilde{r}_{\text{isco}})^3) + \cdots.
\end{align*}
And therefore we have the Ori-Thorne transition regime governing ODE:
\begin{align*}
    \ddot{\tilde{r}}=-\frac{1}{2}\frac{\partial V_{eff}(\tilde{\tau},\tilde{r})}{\partial \tilde{r}}=-\frac{1}{2}\frac{\partial^2 V_{\text{eff}}}{\partial \tilde{\tau} \partial \tilde{r}}(0, \tilde{r}_{\text{isco}})\cdot\tilde{\tau}
- \frac{1}{4} \frac{\partial^3 V_{\text{eff}}}{\partial \tilde{r}^3}(0, \tilde{r}_{\text{isco}})\cdot(\tilde{r} - \tilde{r}_{\text{isco}})^2.
\end{align*}

To be consistent with Ori-Thorne's notation, we first unpack $\frac{\partial^2 V_{\text{eff}}}{\partial \tilde{\tau} \partial \tilde{r}}(0, \tilde{r}_{\text{isco}})$:
\begin{align*}
    \frac{\partial^2 V_{\text{eff}}}{\partial \tilde{\tau} \partial \tilde{r}}(0, \tilde{r}_{\text{isco}})= \frac{\partial}{\partial \tilde{r}}(\frac{\partial V_{\text{eff}}}{\partial \tilde{\tau}})|_{\text{isco}}=\frac{\partial}{\partial \tilde{r}}\left(\frac{\partial V_{\text{eff}}}{\partial \tilde{E}}\frac{\text{d}\tilde{E}}{\text{d}\tilde{\tau}}+\frac{\partial V_{\text{eff}}}{\partial \tilde{L}}\frac{\text{d}\tilde{L}}{\text{d}\tilde{\tau}}\right)|_{\text{isco}}.
\end{align*}
Where \ref{eq:OT transition E lost rate},\ref{eq:OT transition L lost rate} $\Longrightarrow$
\begin{align*}
     \frac{\text{d}\tilde{E}}{\text{d}\tilde{\tau}}|_{\text{isco}}&=-\kappa\eta\tilde{\Omega}_{isco}\\
     \frac{\text{d}\tilde{L}}{\text{d}\tilde{\tau}}|_{\text{isco}}&=-\kappa\eta
\end{align*}
$\Longrightarrow$
\begin{align*}
    \frac{\partial^2 V_{\text{eff}}}{\partial \tilde{\tau} \partial \tilde{r}}(0, \tilde{r}_{\text{isco}})=-\kappa\eta(\frac{\partial^2 V_{\text{eff}}}{\partial \tilde{E} \partial \tilde{r}}(0, \tilde{r}_{\text{isco}})\tilde{\Omega}_{isco}+\frac{\partial^2 V_{\text{eff}}}{\partial \tilde{L} \partial \tilde{r}}(0, \tilde{r}_{\text{isco}})).
\end{align*}
Now define:
\begin{align*}
    \alpha &:=  \frac{1}{4} \frac{\partial^3 V_{\text{eff}}}{\partial \tilde{r}^3}(0, \tilde{r}_{\text{isco}}),\\
    \beta &:=-\frac{1}{2}(\frac{\partial^2 V_{\text{eff}}}{\partial \tilde{E} \partial \tilde{r}}(0,\tilde{r}_{\text{isco}})\tilde{\Omega}_{isco}+\frac{\partial^2 V_{\text{eff}}}{\partial \tilde{L} \partial \tilde{r}}(0, \tilde{r}_{\text{isco}})),\\
    R&:=\tilde{r}-\tilde{r}_{\text{isco}}.
\end{align*}
And our ODE becomes:
\begin{equation}
    \ddot{R} \equiv \frac{\text{d}^2R}{\text{d}\tilde{\tau}^2} = -\alpha R^2 - \kappa\eta\beta\cdot\tilde{\tau} .
\end{equation}
\begin{figure}
\centering
\includegraphics[width=0.7\linewidth]{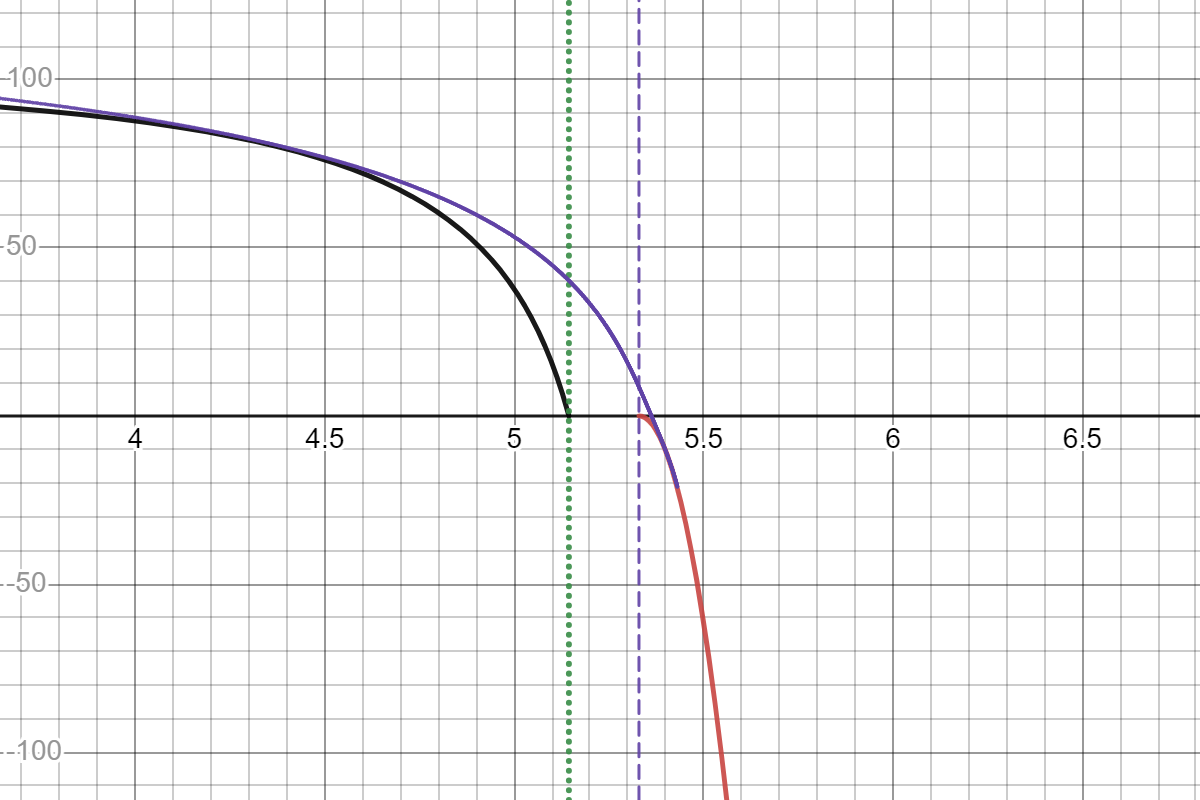}
\caption{\label{fig:OTTransition}Ori-Thorne Transition Procedure\cite{DesmosGraph} The blue dotted line is the ISCO radius, green dotted line is ISCO minus $\delta$ where $\delta$ is the `inward push'. The purple line is Ori-Thorne transition plot on the previous plots of Adiabatic and Plunge. The plot is the dimensionless proper time $\tilde{\tau}$ as function of dimensionless radius $\tilde{r}$. As $\tilde{r}$ getting close to $\tilde{r}_{isco}$, Ori-Thorne transition function takes over from adiabatic to plunge, jumping through the radius where the other two fails. }
\end{figure}
\subsubsection{Numerical solution and dimensionless motion notations}
The ODE we are going to solve is Painlevé equation of first kind. Before we do so we use further simplified notation by setting some constants:
\begin{align*}
    R_0:=(\beta\kappa)^{\frac{2}{5}}\alpha^{-\frac{3}{5}}, \tau_0:=(\alpha\beta\kappa)^{-\frac{1}{5}}.
\end{align*}
And define Ori-Thorne dimensionless notation by re-scaling $\tilde{r},\tilde{\tau}$ c.f. \cite{Ori_2000}:
\begin{equation}
       X:=\frac{R}{R_0}\eta^{-\frac{2}{5}}=\frac{1}{R_0\eta^{\frac{2}{5}}}(\tilde{r}-\tilde{r}_{\text{isco}}),\:T:=\frac{\eta^\frac{1}{5}}{\tau_0}\tilde{\tau},
    \label{eq: OT Dimensionless notation} 
\end{equation}
we will have:
\begin{align*}
    R&=(\beta\kappa\eta)^{\frac{2}{5}}\alpha^{-\frac{3}{5}}X;\\
    \tilde{\tau}&=(\alpha\beta\kappa\eta)^{-\frac{1}{5}}T;\\
    \frac{\text{d}}{\text{d}\tilde{\tau}}&=(\alpha\beta\kappa\eta)^{\frac{1}{5}}\frac{\text{d}}{\text{d}T};\\
    \frac{\text{d}R}{\text{d}\tilde{\tau}}&=(\beta\kappa\eta)^{\frac{3}{5}}\alpha^{-\frac{2}{5}}\frac{\text{d}X}{\text{d}T};\\
    \frac{\text{d}^2R}{\text{d}\tilde{\tau}^2}&=(\beta\kappa\eta)^{\frac{4}{5}}\alpha^{-\frac{1}{5}}\frac{\text{d}^2X}{\text{d}T^2}.
\end{align*}
Thus the Ori-Thorne Transition procedure governing ODE becomes:
\begin{equation}
        \frac{\text{d}^2X}{\text{d}T^2} = -X^2 - T .
        \label{eq:OT second order}
\end{equation}

This is the re-scaled Painlevé equation of the first kind (P-I). Unlike linear differential equations, there are no simple formulas for the solutions of P-I. The solutions to P-I cannot be expressed in terms of elementary functions or even in terms of classical special functions. They are new types of transcendental functions. The general approach to solve the Painlevé I equation, or to study its solutions, involves numerical methods, series solutions, asymptotic analysis etc. Here we are going to use numerical approximation given the difficulty of finding explicit solutions.

See \ref{appendix: OT Numerical Approximation} for detailed calculation of numerical methods. The figure \ref{fig:OTTransition} shows the numerical solution of Ori-Thorne transition regime.

\subsection{Ori-Thorne-Kesden Transition Procedure}
In 2011, Kesden claimed that the assumption of energy-angular momentum relation equation \ref{eq:OT Transition regime E-L relation} leads to mathematical inconsistency\cite{PhysRevD.83.104011}. And thus he relaxed this assumption and added a correction constant to Ori-Thorne Transition ODE

We will look at this by carrying Kesden's derivation forward, and see where the inconsistency comes from and if it is actually an inconsistency.

\subsubsection{Governing equation with relaxed assumption}

Instead of assuming the energy-angular momentum assumption in the Ori-Thorne transition assumptions, Kesden relaxed this assumption and Taylor expanded the radial governing equation \ref{eq: r governing equation SSCD} with unknown energy-angular momentum relation. 

First define re-scaled energy and angular momentum :
\begin{align*}
    \chi&:=\tilde{\Omega}_{isco}^{-1}(\tilde{E}-\tilde{E}_{isco}),\\
    \xi&:=(\tilde{L}-\tilde{L}_{isco}),
\end{align*}
and now we can use $\chi-\xi$ to represent or capture the energy-angular momentum difference.

Taylor expanding equation \ref{eq: r governing equation SSCD} around $\tilde{r}_{isco},\tilde{E}_{isco},\tilde{L}_{isco}$  up to third order in $R\equiv(\tilde{r}-\tilde{r}_{isco})$ and first order in $(\tilde{E}-\tilde{E}_{isco})$ and $(\tilde{L}-\tilde{L}_{isco})$ gives:
\begin{equation}
     \dot{\tilde{r}}^2\equiv \dot{R}^2 \equiv (\frac{\text{d}R}{\text{d}\tilde{\tau}})^2 \simeq -\frac{2}{3}\alpha R^3 +2\beta R\xi+\frac{\partial V_{\text{eff}}}{\partial \tilde{L}}(\chi-\xi)-\tilde{\Omega}\frac{\partial^2 V_{\text{eff}}}{\partial \tilde{E}\partial \tilde{r}}(\chi-\xi)R.
     \label{eq: OTK first order}
\end{equation}
And Kesden found that by assuming $\tilde{E}$ and $\tilde{L}$ as constants, i.e. $\chi,\xi$ are constants and differentiate on both sides we therefore have:
\begin{align*}
    \frac{\text{d}^2R}{\text{d}\tilde{\tau}^2} = -\alpha R^2 +\beta\xi-\frac{1}{2}\tilde{\Omega}\frac{\partial^2 V_{\text{eff}}}{\partial \tilde{E}\partial \tilde{r}}(\chi-\xi).
\end{align*}
Now if we substitute in Ori-Thorne's E-L relation assumption
\begin{align*}
    \chi=\xi=-\kappa\eta\tilde{\tau},
\end{align*}
this can be reduced straight away to Ori-Thorne's transition procedure ODE:
\begin{align*}
   \frac{\text{d}^2R}{\text{d}\tilde{\tau}^2} = -\alpha R^2 -\kappa\eta\beta\tilde{\tau}.
\end{align*}
and switching to Ori-Thorne's dimensionless motion notation \ref{eq: OT Dimensionless notation} this is:
\begin{align*}
    \frac{\text{d}^2X}{\text{d}T^2} = -X^2 - T.
\end{align*}
However if we substitute in Ori-Thorne's E-L relation in equation\ref{eq: OTK first order} we would obtain:
\begin{equation}
    (\frac{\text{d}X}{\text{d}T})^2 = -\frac{2}{3}X^3 - 2XT.
    \label{eq: OTK second order without Y}
\end{equation}

And Kesden found that the Ori-Thorne's second order ODE is not consistency with this. This was shown in a numerical solution plot (Fig 2, \cite{PhysRevD.83.104011}). Also this can be shown by differentiating both sides of above equation and obtains:
\begin{align*}
        \frac{\text{d}^2X}{\text{d}T^2} = -X^2 - T -{X}/{\frac{\text{d}X}{\text{d}T}}.
\end{align*}
Where the extra terms indicating this equation differs with Ori-Thorne's transition procedure ODE.
\begin{figure}
\centering
\includegraphics[width=0.7\linewidth]{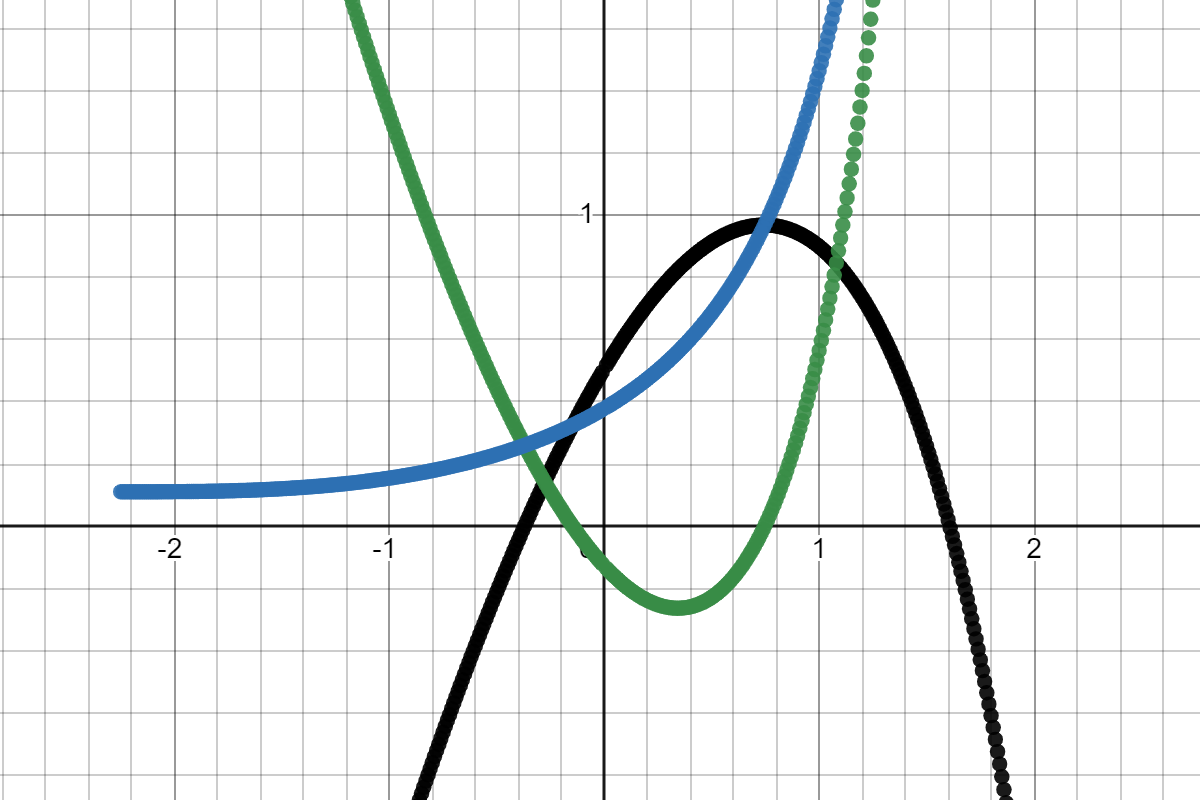}
\caption{\label{fig: Kesden-Y-Correction}Kesden's Y Correction\cite{DesmosGraph} The blue dotted line is left hand side of \ref{eq: OTK second order without Y} with numerical solution of Ori-Thorne's Transition ODE. The green dotted line is the numerical solution of \ref{eq: OTK second order without Y}. There clearly shows a difference between those two, and the difference is plotted as dotted black line. Which dotted black also represents the Kesden's Y correction to Ori-Thorne's Transiton ODE\cite{PhysRevD.83.104011} }
\end{figure}
\subsubsection{Energy - Angular Momentum(E-L) relation and Kesden's Y correction}

Kesden in his paper claims that this extra term can be absorbed by distinguish $\chi$ and $\xi$. By doing so there will be a new term $(\chi-\xi)$ in Ori-Thorne transition ODE. Here set the difference:
\begin{align*}
    \chi-\xi&\equiv\eta^\frac{6}{5}(\chi-\xi)_0Y,\\
    (\chi-\xi)_0&:=\alpha^{-\frac{4}{5}}(\beta\kappa)^\frac{6}{5}(\frac{\partial V_{\text{eff}}}{\partial \tilde{L}})^{-1},
\end{align*}
and substituting in \ref{eq: OTK first order} gives:
\begin{align*}
    (\frac{\text{d}X}{\text{d}T})^2 = -\frac{2}{3}X^3 - 2XT +Y.
\end{align*}
Where this $Y$ term is the extra term resulted, by distinguish $\chi$ from $\xi$. Differentiating it now on both sides w.r.t. $T$ would yield:
\begin{align*}
    \frac{\text{d}^2X}{\text{d}T^2} = -X^2 - T -{X}/{\frac{\text{d}X}{\text{d}T}}+\frac{1}{2}\frac{\text{d}Y}{\text{d}T}/\frac{\text{d}X}{\text{d}T}.
\end{align*}
Kesden then set $\frac{\text{d}Y}{\text{d}T}=2X$ and solves a new E-L relation, thus this new E-L relation will allow equation \ref{eq: OTK first order} to be consistent with Ori-Thorne's transition procedure ODE.  

In later chapters we will approach this from a different view for explaining the inconsistency here. And the reason that we spent a chapter introducing Kesden's view of inconsistency is that the idea of capture inconsistency into E-L relation is reliable, we will seek to extend the Y-correction to absorb `time-dependent radial background force'.

\subsection{Time-dependent Radial Background Force Extended Transition Procedure}
In Ori-Thorne's approach of transition regime, the time-dependent radial background force is being neglected (3.10 \cite{Ori_2000}).  We will look at adding radial self force back into Ori-Thorne's transition ODE, later we will see that this can explain the inconsistency between Ori-Thorne and Kesden. Then we will extend the model with time-dependent radial background force added.

\subsubsection{Radial Taylor Expansion around ISCO}

We go back to equation \ref{eq: r governing equation SSCD} and start over from here:
\begin{align*}
    \dot{\tilde{r}}^2=f(\tilde{E},\tilde{L},\tilde{r})=\tilde{E}^2-1+\frac{2}{\tilde{r}}-\frac{\tilde{L}^2-\tilde{a}^2\tilde{E}^2+\tilde{a}^2}{\tilde{r}^2}+\frac{2(\tilde{L}-\tilde{a}\tilde{E})^2}{\tilde{r}^3}.
\end{align*}
We use all the same assumptions as Ori-Thorne did as in \ref{Chapter: OT Approximation}, which means all the facts (\ref{eq:Veff expansion}) we used in Taylor expanding $V_{eff}(\tilde{r})$ are still valid. And the E-L relation (\ref{eq:OT Transition regime E-L relation}) is still the same thus energy and angular momentum lost rate are just equation \ref{eq:OT transition E lost rate},\ref{eq:OT transition L lost rate}.
Taylor expanding $f(\tilde{\tau},\tilde{r})$ instead of $V_{eff}(\tilde{\tau},\tilde{r})$ up to the third order of $\tilde{\tau},\tilde{r}$ with
\begin{align*}
    f(\tilde{\tau},\tilde{r})\equiv\tilde{E}(\tilde{\tau})^2-1-V_{eff}(\tilde{\tau},\tilde{r}):
\end{align*}
\begin{align*}
    f(\tilde{\tau},\tilde{r})&= \frac{\partial \tilde{E}^2}{\partial \tilde{\tau}}(0, \tilde{r}_{\text{isco}})\cdot\tilde{\tau}  - \frac{\partial V_{\text{eff}}}{\partial \tilde{\tau}}(0, \tilde{r}_{\text{isco}})\tilde{\tau}  \\&- \frac{1}{2!}\left(-\frac{\partial^2 \tilde{E}^2}{\partial \tilde{\tau}^2}(0, \tilde{r}_{\text{isco}})\cdot\tilde{\tau}^2+\frac{\partial^2 V_{\text{eff}}}{\partial \tilde{\tau}^2}(0, \tilde{r}_{\text{isco}})\cdot\tilde{\tau}^2 + 2\frac{\partial^2 V_{\text{eff}}}{\partial \tilde{\tau} \partial \tilde{r}}(0, \tilde{r}_{\text{isco}})\cdot\tilde{\tau}(\tilde{r} - \tilde{r}_{\text{isco}}) \right)
\\&- \frac{1}{3!}( \frac{\partial^3 V_{\text{eff}}}{\partial \tilde{r}^3}(0, \tilde{r}_{\text{isco}})\cdot(\tilde{r} - \tilde{r}_{\text{isco}})^3) + \cdots.
\end{align*} 
Similarly using simplification as we did in \ref{eq:Veff expansion} we will have:
\begin{align}
    \dot{\tilde{r}}^2= \frac{\partial f}{\partial \tilde{\tau}}(0, \tilde{r}_{\text{isco}})\cdot\tilde{\tau}+\frac{1}{2}\frac{\partial^2 f}{\partial \tilde{\tau}^2}(0, \tilde{r}_{\text{isco}})\cdot\tilde{\tau}^2-\frac{\partial^2 V_{\text{eff}}}{\partial \tilde{\tau} \partial \tilde{r}}(0, \tilde{r}_{\text{isco}})\cdot\tilde{\tau}R-\frac{1}{6}\frac{\partial^3 V_{\text{eff}}}{\partial \tilde{r}^3}(0, \tilde{r}_{\text{isco}})\cdot R^3.
\end{align}

Where the third and fourth term coefficients have been well defined in Ori-Thorne's notation with $\alpha$ and $\beta$, we will try to unpack the first and second term here and seek coefficient for those to simplify calculations. The reason we're not simply replacing the first and second term by constants using some symbol is that it is important to unravel and track the relation between the coefficients and some important constants, $\eta$ for example. Unpacking the first term we can see that:
\begin{align*}
    \frac{\partial f}{\partial \tilde{\tau}}(0, \tilde{r}_{\text{isco}})&=\frac{\partial f}{\partial \tilde{E}}\frac{\partial \tilde{E}}{\partial \tilde{\tau}}(0, \tilde{r}_{\text{isco}})+\frac{\partial f}{\partial \tilde{L}}\frac{\partial \tilde{L}}{\partial \tilde{\tau}}(0, \tilde{r}_{\text{isco}})\\
    &=-\tilde{\Omega}_{isco}\kappa\eta\frac{\partial f}{\partial \tilde{E}}(0, \tilde{r}_{\text{isco}})-\kappa\eta\frac{\partial f}{\partial \tilde{L}}(0, \tilde{r}_{\text{isco}}).
\end{align*}
Where
\begin{align*}
    \tilde{\Omega}_{isco}\frac{\partial f}{\partial \tilde{E}}(0, \tilde{r}_{\text{isco}})+\frac{\partial f}{\partial \tilde{L}}(0, \tilde{r}_{\text{isco}})\equiv(\tilde{\Omega}\frac{\partial f}{\partial \tilde{E}}+\frac{\partial f}{\partial \tilde{L}})|_{isco}.
\end{align*}
only contains constants $\tilde{r}_{isco},\tilde{E}_{isco},\tilde{L}_{isco}$. And we define this constants by $-\gamma$ and obtains first term:
\begin{align*}
    \frac{\partial f}{\partial \tilde{\tau}}(0, \tilde{r}_{\text{isco}})=  \kappa\eta \gamma,\\
    \gamma:=-[\tilde{\Omega}\frac{\partial f}{\partial \tilde{E}}+\frac{\partial f}{\partial \tilde{L}}]|_\text{isco}.
\end{align*}
For the second term, using the fact that 
\begin{align*}
    f\equiv(\kappa\eta)^2(\tilde{\Omega}_{isco}^2-\frac{1}{\tilde{r}^2}-\frac{\tilde{a}^2\tilde{\Omega}_{isco}^2}{\tilde{r}^2}+\frac{2}{\tilde{r}^3}+\frac{2\tilde{a}^2\tilde{\Omega}_{isco}^2}{\tilde{r}^3}-\frac{4\tilde{a}\tilde{\Omega}_{isco}}{\tilde{r}^3})\tilde{\tau}^2,
\end{align*}
we have:
\begin{align*}
    \frac{1}{2}\frac{\partial^2 f}{\partial \tilde{\tau}^2}&(0, \tilde{r}_{\text{isco}})=(\kappa\eta)^2\sigma,\\
    \sigma:=[\tilde{\Omega}^2-\frac{1}{\tilde{r}^2}-&\frac{\tilde{a}^2\tilde{\Omega}^2}{\tilde{r}^2}+\frac{2}{\tilde{r}^3}+\frac{2\tilde{a}^2\tilde{\Omega}^2}{\tilde{r}^3}-\frac{4\tilde{a}\tilde{\Omega}}{\tilde{r}^3}]|_{isco}.
\end{align*}
Here $\sigma$  is just some constant. And together using $\alpha$ and $\beta$ from \ref{eq:Veff expansion}$\Longrightarrow$
\begin{align*}
    -\frac{\partial^2 V_{\text{eff}}}{\partial \tilde{\tau} \partial \tilde{r}}(0, \tilde{r}_{\text{isco}})=-2\beta\kappa\eta,\\
    -\frac{1}{6}\frac{\partial^3 V_{\text{eff}}}{\partial \tilde{r}^3}(0, \tilde{r}_{\text{isco}})=-\frac{2}{3}\alpha,
\end{align*}
therefore we have:
\begin{equation}
    \dot{R}^2\equiv\dot{\tilde{r}}^2=f=\kappa\eta\gamma\tilde{\tau}+(\kappa\eta)^2\sigma\tilde{\tau}^2-2\beta\kappa\eta\tilde{\tau}R-\frac{2}{3}\alpha R^3.
    \label{eq: SE transition first order}
\end{equation}
Under Ori-Thorne dimensionless motion notation, this is:
\begin{equation}
    (\frac{\text{d}X}{\text{d}T})^2=-\frac{2}{3}X^3-2XT+\tilde{\gamma}\eta^{-\frac{2}{5}}T+\tilde{\sigma}\eta^{\frac{2}{5}}T^2,
    \label{eq: SE transition first order dimensionless}
\end{equation}
\begin{align*}
    \tilde{\gamma}&:=\beta^{-\frac{7}{5}}\kappa^{-\frac{2}{5}}\alpha^{\frac{3}{5}}\gamma,\\
    \tilde{\sigma}&:=\beta^{-\frac{8}{5}}\kappa^{\frac{2}{5}}\alpha^{\frac{2}{5}}\sigma.
\end{align*}

\subsubsection{Time-dependent Radial Background Force}
In Ori-Thorne's derivation of radial governing equation, there is a step that this time-dependent force is neglected. We shall take a close look at this step. From equation \ref{eq: SE transition first order}, by differentiating both sides we will have:
\begin{align*}
     \ddot{R}=\frac{1}{2\dot{R}}\frac{\partial f}{\partial \tilde{\tau}}+\frac{1}{2}\frac{\partial f}{\partial R}.
\end{align*}
As stated in \ref{eq:Veff expansion}, the $\frac{\partial f}{\partial \tilde{\tau}}$ is related to time-dependent radial background force, thus the first term got neglected. Take a more closer look and unpack these terms we will get:
\begin{align*}
    \frac{1}{2\dot{R}}\frac{\partial f}{\partial \tilde{\tau}}&=\frac{1}{2\dot{R}}(\kappa\eta\gamma+2(\kappa\eta)^2\sigma\tilde{\tau}-2\beta\kappa\eta R),\\
    \frac{1}{2}\frac{\partial f}{\partial R}&=-\beta\kappa\eta\tilde{\tau}-\alpha R^2.
\end{align*}
Again if we neglect the self force term we will recover Ori-Thorne's Transition ODE. We know that $\tilde{\tau}\propto \eta^{-1/5}$, $R\propto\eta^{2/5}$ and $\frac{1}{\dot{R}}\propto\eta^{-3/5}$ from the Ori-Thorne's dimensionless notation settings. Thus this time-dependent radial background force component has leading order of $\eta^{6/5}$ and to the smallest leading order of $\eta^{2/5}$. Thus this suggest that we need to be extremely careful when dealing with this term. As $\eta$ goes to 0 it could be a potential problem as $\eta^{2/5}$ blows up quicker than $\ddot{R}\propto\eta^{4/5}$. Now we switch to Ori-Thorne's dimensionless motion notation to gain a better view, by either switching this second order ODE or straight away differentiate equation \ref{eq: SE transition first order dimensionless} on both sides we will have:
\begin{equation}
    \frac{\text{d}^2X}{\text{d}T^2}=-X/\frac{\text{d}X}{\text{d}T}+\frac{\tilde{\gamma}\eta^{-\frac{2}{5}}}{2}/\frac{\text{d}X}{\text{d}T}+\tilde{\sigma}\eta^{\frac{2}{5}}/\frac{\text{d}X}{\text{d}T}\mathbf{-T-X^2}.
    \label{eq: SE transition second order dimensionless}
\end{equation}
The un-bold terms comes from time-dependent radial background force component. It becomes clear now that the term, $-X/\frac{\text{d}X}{\text{d}T}$ actually comes from the this component, which has been neglected during derivation from Ori-Thorne, however it was picked up in Kesden. We have shown here that there is no fundamental mathematical inconsistency during Ori-Thorne's derivation and there is no first-order ODE in Ori-Thorne's transition procedure that can straight away lead to the second-order ODE equation \ref{eq:OT second order}, as the term $-XT$ contains both time-dependent and time-independent background force. The Kesden Y correction is in fact targeting at cancelling out the first term of this time-dependent background force. 

\subsubsection{Coupling the time-dependent background force with Y correction}

Since this time-dependent background force is the force due to energy-angular momentum loss, coupling this force into the difference of energy-angular momentum relation could be an good explanation of this. As the energy lost and angular momentum lost ratio should also be depend on the test particle's mass if it does not travels on a perfect-circular orbit during transition regime.

And we can do this following Kesden's Y correction, $Y_{k}$. We have known that:
\begin{align*}
    \frac{\text{d}Y_k}{\text{d}T}=2X.
\end{align*}
And now we shall look for some modified $\tilde{Y}$ such that
\begin{align*}
   (\frac{\text{d}X}{\text{d}T})^2=-\frac{2}{3}X^3-2XT+\tilde{\gamma}\eta^{-\frac{2}{5}}T+\tilde{\sigma}\eta^{\frac{2}{5}}T^2+\tilde{Y}
\end{align*}
is consistent with equation \ref{eq: SE transition second order dimensionless}.
And this can be accomplished by setting:
\begin{equation}
    \tilde{Y}(T):=Y_k(T)-\tilde{\gamma}\eta^{-\frac{2}{5}}T-\tilde{\sigma}\eta^{\frac{2}{5}}T^2,
    \label{eq: Y correction}
\end{equation}
and thus:
\begin{align*}
    \frac{\text{d}\tilde{Y}}{\text{d}T}=2X-2\tilde{\sigma}\eta^{\frac{2}{5}}T-\tilde{\gamma}\eta^{-\frac{2}{5}}.
\end{align*}
Now our first order transition ODE couples all forces. 

Further on, in the limit of maximal spin, Kesden found that in order to make Y correction further accurate, Y has to evolve according to:
\begin{align*}
    \frac{\text{d}Y'_k}{\text{d}T}= 2X + 2\eta^{2/5} CY'_k\frac{\text{d}X}{\text{d}T}.
\end{align*}
For some constant $C$, by including an extra correction term from the Taylor expansion. Where this term extra term evolves $\propto \eta^{2/5}$. We can see that from our correction that we arrive at a similar correction term from including the time-dependent background force, which is also $\propto \eta^{2/5}$. This suggests the consistency and validation of our correction and it will be strongly instructive to compare these two terms under high spin.

\subsubsection{Time-dependent Background Force Extended Ori-Thorne Transition}

From our previous calculation we can see that the time-dependent background force could be crucial and non-negligible. Beside interpreting the term inside energy-angular momentum difference ($\tilde{Y}$), we shall also follow Ori-Thorne's derivation and add this self-force term back. Assuming again the energy-angular momentum difference follows Ori-Thorne's assumption, we will have a new first order ODE: 
\begin{align*}
    (\frac{\text{d}X}{\text{d}T})^2=-\frac{2}{3}X^3-2XT+\tilde{\gamma}\eta^{-\frac{2}{5}}T+\tilde{\sigma}\eta^{\frac{2}{5}}T^2.
\end{align*}

\section{Adiabatic inspiral perturbation induced plunge phase}

The transition regime offers a method for interpreting and addressing the fail of the Adiabatic and Plunge regimes near the ISCO. It focuses on refining approximations around the ISCO for Adiabatic inspiral and Plunge regimes to derive a new governing equation that enables a test particle to smoothly transition through the ISCO.

Here, we propose an alternative approach. Instead of merely adjusting certain approximations, we begin with the adiabatic inspiral and introduce perturbations. Given that no test particles travel on perfectly circular orbits, we posit and will justify that even minor perturbations added to the adiabatic process will lead to the particle entering a distinctly different plunge phase. This concept is akin to solving wave equations in Quantum Physics across an interface: despite potential differences that are significant or even discontinuous in theoretical cases, the continuity of the derivative is still required.

\begin{figure}
\centering
\includegraphics[width=0.5\linewidth]{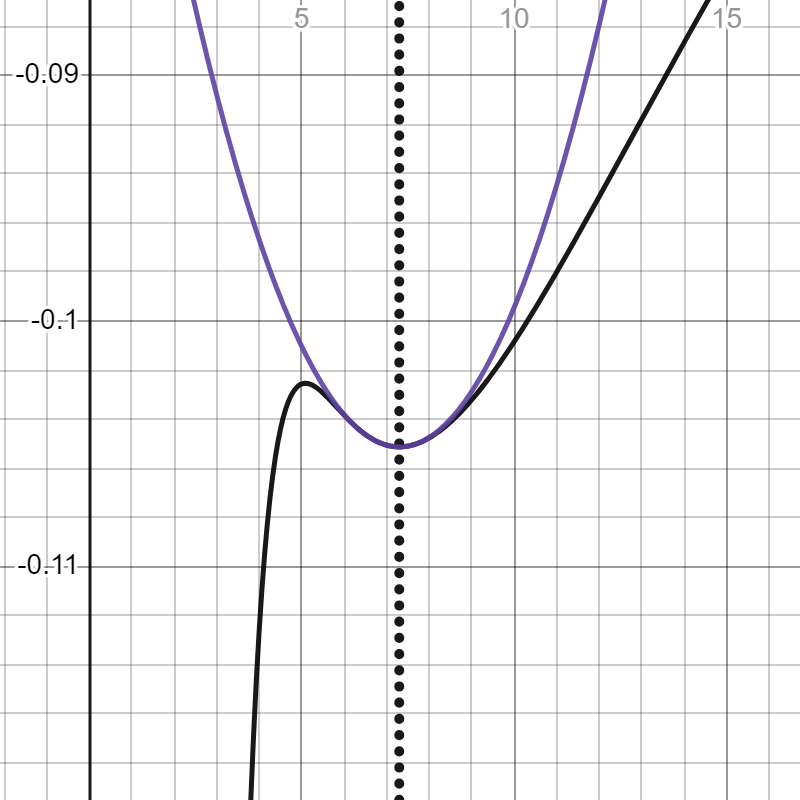}
\caption{\label{fig: Oscillation Coefficient}Oscillator Approximation\cite{DesmosGraph}  The black line is the plot of effective potential, the black dotted line indicates the stable circular orbit radius $\tilde{r}_s$, and the purple line is harmonic oscillation approximation, $\tilde{a}=0$ (Schwarzschild)}
\end{figure}

\subsection{Oscillation frequency coefficient}
We start again using the radial governing equation under effective potential \ref{eq: r governing equation with effective potential SSCD}:
\begin{align*}
    V_{eff}(\tilde{r})=-\frac{2}{\tilde{r}}+\frac{\tilde{L}^2-\tilde{a}^2\tilde{E}^2+\tilde{a}^2}{\tilde{r}^2}-\frac{2(\tilde{L}-\tilde{a}\tilde{E})^2}{\tilde{r}^3}.
\end{align*}
And we previously solved that the minimal and maximal point, corresponding to the unstable circular orbit and stable circular orbit under SSCD notation is:
\begin{align*}
    \tilde{r}_{c\pm}=\frac{\tilde{L}^2-\tilde{a}^2\tilde{E}^2+\tilde{a}^2\pm\sqrt{(\tilde{L}^2-\tilde{a}^2\tilde{E}^2+\tilde{a}^2)^2-12(\tilde{L}-\tilde{a}\tilde{E})^2}}{2}.
\end{align*}
Here the stable circular orbit is at $\tilde{r}_s\equiv\tilde{r}_{c+}$, the unstable circular orbit is at $\tilde{r}_u\equiv\tilde{r}_{c-}$. We then Taylor expand the effective potential around the stable circular orbit, assuming that $E,L$ are constants: 
\begin{align*}
    V_{\text{eff}}(\tilde{r}-\tilde{r}_s) &= V_{\text{eff}}(\tilde{r}_s) + \frac{\partial V_{\text{eff}}}{\partial \tilde{r}}(\tilde{r}_s)\cdot(\tilde{r}-\tilde{r}_s) + \frac{\partial^2 V_{\text{eff}}}{2\partial \tilde{r}^2}(\tilde{r}_s)\cdot(\tilde{r}-\tilde{r}_s)^2+O((\tilde{r}-\tilde{r}_s)^3)\\&=\frac{\partial^2 V_{\text{eff}}}{2\partial \tilde{r}^2}(\tilde{r}_s)\cdot(\tilde{r}-\tilde{r}_s)^2+O((\tilde{r}-\tilde{r}_s)^3)+constant.
\end{align*}
Neglecting the higher order terms, we will have a harmonic oscillator-like motion with adiabatic inspiral oscillation frequency coefficient (or angular frequency if it is harmonic oscillator) is 
\begin{equation}
    \omega^2_{aip}(\tilde{r}_s):=\frac{\partial^2 V_{\text{eff}}}{2\partial \tilde{r}^2}({\tilde{r}_s})=-\frac{2}{\tilde{r}^3_s}+3\frac{\tilde{L}^2_s-\tilde{a}^2\tilde{E}^2_s+\tilde{a}^2}{\tilde{r}^4_s}-12\frac{(\tilde{L}_s-\tilde{a}\tilde{E}_s)^2}{\tilde{r}^5_s}.
    \label{eq: aip coefficient}
\end{equation}

And here we simply assume that $\tilde{E}$ and $\tilde{L}$ are the energy and angular momentum at circular orbit radius $\tilde{r}_s$ using equation \ref{eq:E stable circular orbit SSCD}, \ref{eq:L stable circular orbit SSCD}, as this is just the perturbation upon the circular orbit. Thus the effective potential can be written as:
\begin{align*}
     V_{\text{eff}}(\mathcal{R}) = \omega^2_{aip}(\tilde{r}_s) &\mathcal{R}^2 + O(\mathcal{R}^3 ) + constant,\\
     \mathcal{R}:= \tilde{r}-&\tilde{r}_s\equiv\tilde{r}_p-\tilde{r}_s.
\end{align*}
Where $\tilde{r}_p$ here is our perturbed solution. And we have our harmonic oscillator approximation ODE by substitute this in and differentiate on both sides and obtains:
\begin{equation}
    \ddot{\mathcal{R}}+\omega^2_{aip}(\tilde{r}_s)\mathcal{R}=0.
    \label{eq: aip oscillation ODE}
\end{equation}

Using formulas above and results in previous chapters one can show that 
\begin{align*}
    \omega^2_{aip}(\tilde{r}_s&)>0\text{  for  }\tilde{r}_s>\tilde{r}_{isco},\\
    &\omega^2_{aip}(\tilde{r}_{isco})=0.
\end{align*}

\subsection{Minimal oscillation radius}

\begin{figure}
\centering
\includegraphics[width=0.65\linewidth]{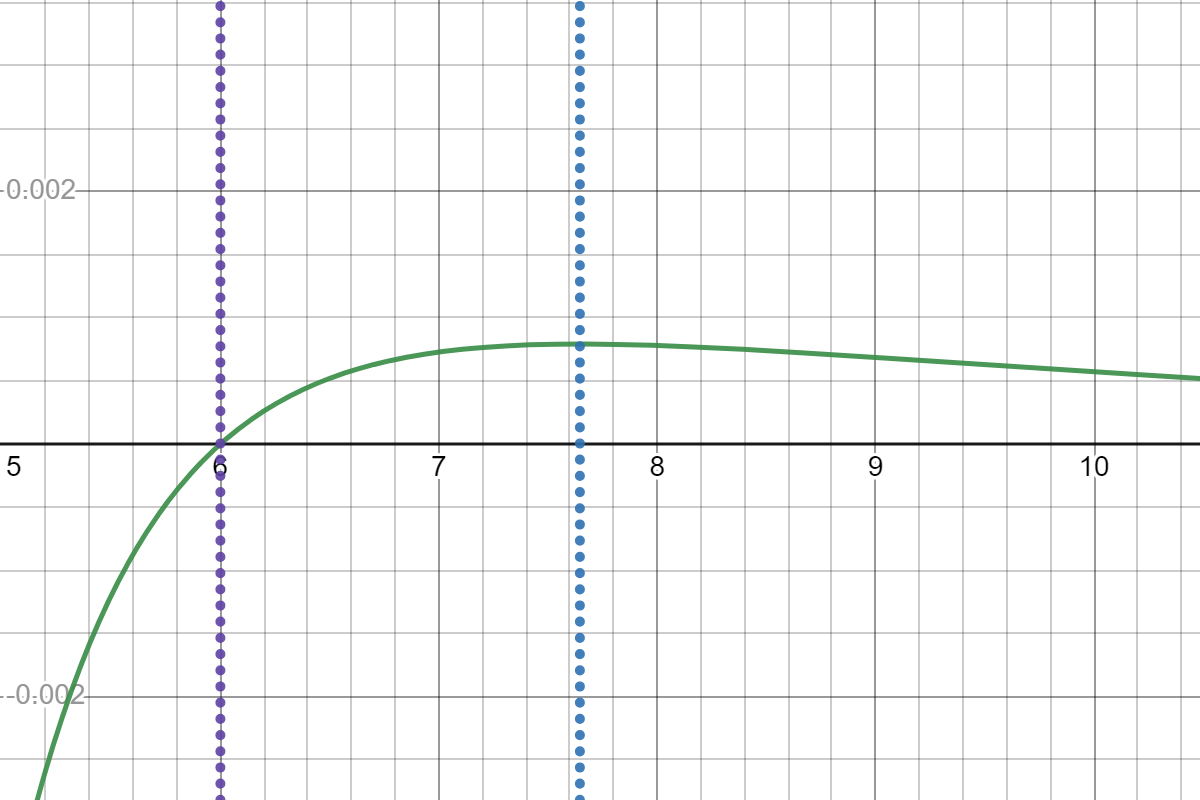}
\caption{\label{fig: r_aimp}Oscillation coefficient against SCO radius\cite{DesmosGraph}  The green line is the plot of Oscillation Coefficient $\omega^2_\text{aip}(\tilde{r}_s)$, the purple dotted line indicates the ISCO, and the blue dotted line is the minimal oscillation radius(Most Stable Circular Orbit) $\tilde{r}_\text{msco}$, $\tilde{a}=0$ (Schwarzschild)}
\end{figure}

We have worked out the oscillation coefficient $\omega^2_{aip}$ now, which is fundamentally the leading order behaviour of an test particle that slightly perturbed from adiabatic inspiral period, just like many other effective potentials analysis in physics. Ideally if we have a constant coefficient then this problem is settled, as the perturbed motion will simply be a harmonic epicycle motion on top of adiabatic inspiral motion.

Here we can see from the plot of the oscillation coefficient $\omega^2_\text{aip}$ that the oscillator coefficient changes as radius evolves in adiabatic inpiral. In simpler words this means when the particle is at different stable orbit, the oscillation frequency and amplitude will change. The coefficient is well above 0 as we calculated previously, and the reason that it is 0 at ISCO is also talked about when we identify the ISCO as innermost circular orbit which is a saddle point thus second order term must vanish. We shall neglect the part smaller than ISCO as we are talking about adiabatic inspiral, but this is also intuitive since there simply won't be any stable orbit after ISCO, thus second order term in effective potential must not be greater or equal 0. 

Moreover, we notice that the coefficient is not linearly decrease as radius decrease, it attains its maximum value at some radius. This is a fun fact that could be used verified the correctness of Adiabatic Inspiral Regime that we will discuss later.  Since the maximum oscillation frequency coefficient indicates maximal frequency and minimal amplitude, we will call it the minimal oscillation coefficient and call the radius Adiabatic Inspiral Minimal Perturbation Orbit $\tilde{r}_\text{msco}$ here, since it is also the radius of Most Stable Circular Orbit, again we will discuss more about this later.

The next step is to work out this radius. The working is straight forward as:
\begin{align*}
    \frac{\partial(\omega^2_{aip}(\tilde{r}_s))}{\partial \tilde{r}_s}|_{\tilde{r}_s=\tilde{r}_\text{msco}}=0,
\end{align*}
and the closed form solution of this radius is complicated. We will look at its numerical solutions and equations for now. If we unpack this and the full equations that solves for $\tilde{r}_{msco}$ is as following:
\begin{align*}
    &[r^{-0.5}(r^{0.5}\left(14.0a^2r^{2.5} + 20ar^4 - 48ar^3 + 40.5r^{3.5} - 33.0r^{4.5} + 6.5r^{5.5}\right)\\
    &\times\left(6a^3 - 25a^2r^{0.5} + 3a^2r^{1.5} - 10ar^2 + 36ar - 18r^{1.5} + 9r^{2.5} - r^{3.5}\right) \\
    &+\left(12.5a^2 + r^{0.5}\left(-4.5a^2r^{0.5} + 20ar - 36a + 27.0r^{0.5} - 22.5r^{1.5} + 3.5r^{2.5}\right)\right)\\&
    \times\left(4a^2r^{3.5} + 4ar^5 - 12ar^4 + 9r^{4.5} - 6r^{5.5} + r^{6.5}\right))]\\
    &\div \left[{\left(4a^2r^{3.5} + 4ar^5 - 12ar^4 + 9r^{4.5} - 6r^{5.5} + r^{6.5}\right)^2}\right]\\
    &=0,
\end{align*}

where $r$ stands for $\tilde{r}_{msco}$ and $a$ stands for $\tilde{a}\equiv J/M^2$. The numerical solutions are listed in Table \ref{tab: aimp}.
\begin{table}
\centering
\begin{tabular}{l|rll}
$\tilde{a}$& $\tilde{r}_\text{isco}$& $\tilde{r}_\text{msco}$&$\omega^2_{aimp}$\\\hline
-0.99& 8.972& 11.465&0.000242\\
 -0.9& 8.717& 11.138&0.000263\\
 -0.5& 7.555& 9.643&0.000402\\
0& 6.000& 7.646&0.000793\\
 0.2& 5.329& 6.784&0.001123\\
 0.5& 4.233& 5.376&0.002205\\
 0.8& 2.907& 3.673&0.006554\\
 0.9& 2.321& 2.920&0.012432\\
 0.99& 1.454& 1.806&0.04335\\
 0.999& 1.182& 1.467&0.06662\\\end{tabular}
\caption{\label{tab: aimp}Innermost stable circular orbit radius; Adiabatic inspiral minimal perturbation radius(Most Stable Circular Orbit) and Adiabatic inspiral minimal perturbation angular frequency}
\end{table}
We can see that the Adiabatic Inspiral Minimal Perturbation Orbit Radius is a constant only depend on $\tilde{a}$ under SSCD notations. This is also a intrinsic property of Kerr Black Hole that is independent of properties of test particles. And for Schwarzschild black hole, the original radius would be:
\begin{align*}
    r_{msco}\simeq\frac{7.646GM}{c^2}.
\end{align*}

In fact, this Adiabatic Inspiral Minimal Perturbation Orbit is also the Most Stable Circular Orbit(MSCO). This is due to the procedure of solving adiabatic inspiral minimal perturbation is mathematically identical for solving the circular orbit with most stability. Since the Adiabatic Inspiral Minimal Perturbation Orbit Radius, or Most Stable Circular Orbit is a intrinsic property of black hole, it should be somewhat detectable or verifiable using observations. Despite any starting initial perturbation of the test particle that travels under near-adiabatic inspiral regime, it should appear to be minimal deviating from exact adiabatic inspiral solution exactly at this radius. 

For observations, we would expect the observation data , for example from Laser Interferometer Space Antenna (LISA) project, that the observed test particle circular trajectories clustering around MSCO radius. That is the observed gravitational waves from binary Kerr black hole sources would appears a peak around MSCO radius, if the massive test particles are randomly distributed and follows near-circular orbit. As the coefficient is positively correlated to spin parameter $\tilde{a}$, thus this should be more significant for high spin Kerr Black hole and prograde test particles.

\subsection{Adiabatic Oscillation Damping}

The step is to look at how the oscillation coefficient varies, the damping effect is similar to classical physics oscillator damping, but this is due to the intrinsic property of Kerr black hole but not heat loss. Most importantly, the oscillation coefficient goes to 0 as test particles gets closer to ISCO. This means that the oscillation amplitude will blow up near ISCO. We will discuss about this later.

As in Adiabatic Inspiral we already obtain a exact trajectory solution \ref{eq: Adiabatic}, by substituting this in \ref{eq: aip oscillation ODE}  we can arrive at the adiabatic inspiral oscillation ODE:
\begin{align*}
    \ddot{\mathcal{R}}+\omega^2_{aip}(\tilde{r}_s(\tilde{\tau}))\mathcal{R}=0,
\end{align*}
This is a long ODE and we shall seek for some simplification. To be consistent and more importantly comparable to Ori-Thorne Transition Regime Procedure, we will first convert to Ori-Thorne Dimensionless notation. From earlier we know that $\mathcal{R}=\tilde{r}_p-\tilde{r}_s$ and $X = \frac{1}{R_0\eta^\frac{2}{5}}(\tilde{r}_s-\tilde{r}_{isco})$, first we set:
\begin{align*}
    X_p:=\frac{1}{R_0\eta^\frac{2}{5}}(\tilde{r}_p-\tilde{r}_{isco}),\\
    \mathcal{X}:=X_p-X= \frac{\mathcal{R}}{R_0\eta^\frac{2}{5}}.
\end{align*}
Where $X_p$ is our dimensionless perturbed solution. And we then have:
\begin{align*}
    \frac{\text{d}^2}{\text{d}\tilde{\tau}^2}&=(\alpha\beta\kappa\eta)^{\frac{2}{5}}\frac{\text{d}^2}{\text{d}T^2},\\
    \ddot{\mathcal{R}}\equiv\frac{\text{d}^2}{\text{d}\tilde{\tau}^2}&R=(\beta\kappa\eta)^\frac{4}{5}\alpha^{-\frac{1}{5}}\frac{\text{d}^2\mathcal{X}}{\text{d}T^2}.
\end{align*}
\begin{figure}
\centering
\includegraphics[width=0.5\linewidth]{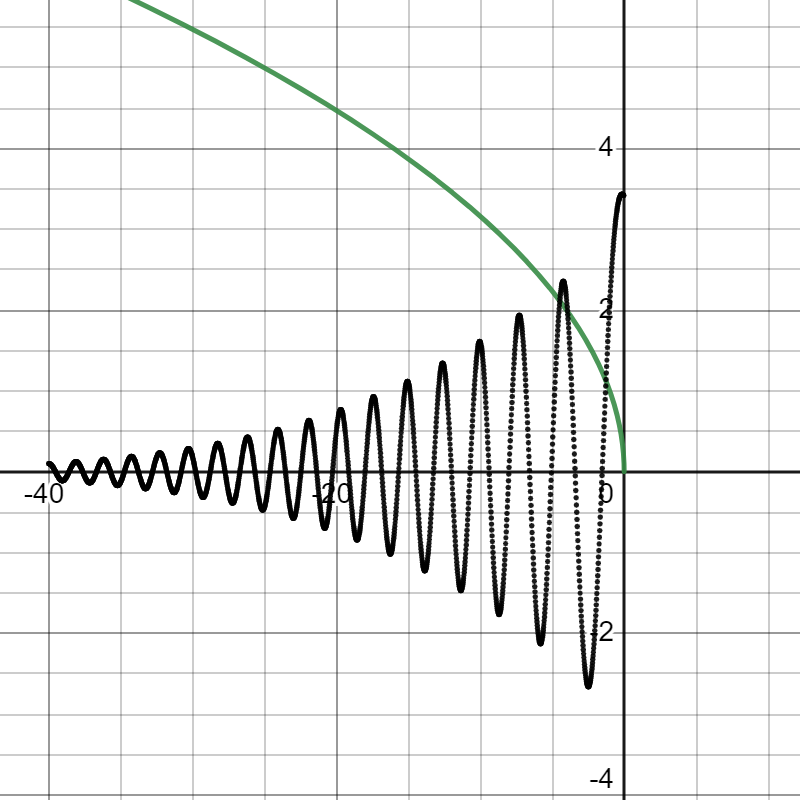}
\caption{\label{fig: aip1}Adiabatic Inspiral Relative Perturbation\cite{DesmosGraph}  The green line is the plot of Adiabatic inpiral $X(T)$, the black dotted line is the plot of adiabatic inspiral perturbation relative to Adiabatic Inspiral $\mathcal{X}(T)$, with $\tilde{a}=0$ (Schwarzschild), $\eta=10^{-5}$}
\includegraphics[width=0.5\linewidth]{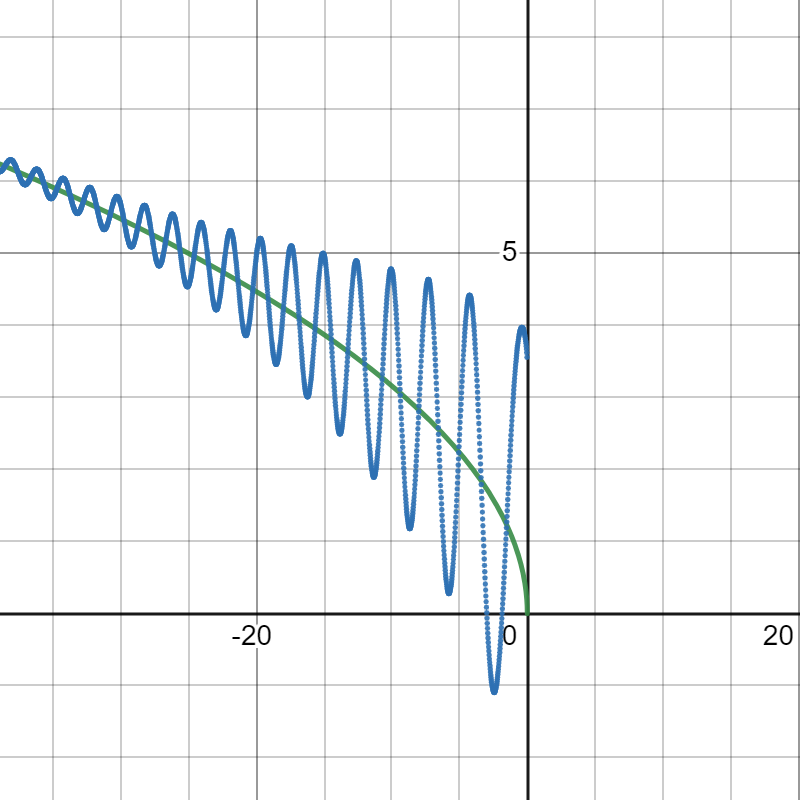}
\caption{\label{fig: aip2}Adiabatic Inspiral Perturbation\cite{DesmosGraph} This is the same plot of previous one but added perturbation onto the adiabatic inspiral. The blue dotted line is perturbed adiabatic inspiral solution $X_p(T)$}
\end{figure}
If we can find some reasonable simplification of $\omega^2_{aip}(\tilde{r}_s)$ then we will be able to obtains a qualitative oscillation ODE to solve. The closest simplification form we find here works well for $\tilde{a}<<1$ , and is accurate at $\tilde{a}=0$. Although it still predicts well for $a\sim1$ but occurs a $\sim10\%$ deviation from explicit results:
\begin{align*}
    \omega^2_{aip}(\tilde{r}_s)\simeq \frac{(\tilde{r}_s-\tilde{r}_{isco})}{\tilde{r}_s^3(\tilde{r}_s-\tilde{r}_{isco}/2)},
\end{align*}
 and oscillating ODE is:
 \begin{align*}
     \ddot{\mathcal{R}}+\frac{(\tilde{r}_s(\tilde{\tau})-\tilde{r}_{isco})}{\tilde{r}_s(\tilde{\tau})^3(\tilde{r}_s(\tilde{\tau})-\tilde{r}_{isco}/2)}\mathcal{R}=0.
 \end{align*}
Switching to Ori-Thorne dimensionless notations, this is:
\begin{align*}
    (\beta\kappa\eta)^\frac{4}{5}\alpha^{-\frac{1}{5}}\frac{\text{d}^2\mathcal{X}}{\text{d}T^2}+(\beta\kappa\eta)^\frac{2}{5}\alpha^{-\frac{3}{5}}\frac{(\beta\kappa\eta)^\frac{2}{5}\alpha^{-\frac{3}{5}}X}{(R_0\eta^\frac{2}{5}X+\tilde{r}_{isco})^3(R_0\eta^\frac{2}{5}X+\tilde{r}_{isco}/2)}\mathcal{X}=0,
\end{align*}
and after simplification this is 
\begin{align*}
    \frac{\text{d}^2\mathcal{X}}{\text{d}T^2}+\frac{X(T)}{\alpha(R_0\eta^\frac{2}{5}X(T)+\tilde{r}_{isco})^3(R_0\eta^\frac{2}{5}X(T)+\tilde{r}_{isco}/2)}\mathcal{X}=0.
\end{align*}
Where from equation 3.23 in \cite{Ori_2000} , adiabatic inspiral near ISCO is $X(T)=\sqrt{-T}$. 

If we take $\eta:=10^{-5}$ and set $\tilde{a}=0$: 
\begin{align*}
    \frac{\sqrt{-T}}{\alpha(R_0\eta^\frac{2}{5}\sqrt{-T}+\tilde{r}_{isco})^3(R_0\eta^\frac{2}{5}\sqrt{-T}+\tilde{r}_{isco}/2)}\sim \nu \sqrt{-T},
\end{align*}
\begin{align*}
    \nu\simeq1.772.
\end{align*}
where $\nu$ is worked out using numerical optimisation for function approximation. And now the ODE under this case is:
\begin{align*}
    \frac{\text{d}^2\mathcal{X}}{\text{d}T^2}+\nu\sqrt{-T}\mathcal{X}=0.
\end{align*}
By solving for $\mathcal{X}(T)$, we will obtain the perturbation solution $X_p(T)=X(T)+\mathcal{X}(T)$ near ISCO. As we previous mentioned, due to the coefficient vanishes the oscillation blows up near ISCO, as we can see in the plot \ref{fig: aip1} and \ref{fig: aip2}. This perturbation even crosses ISCO back and forth.

\subsection{Oscillation Boundary Limit}

Following our previous discussion, there must exist a limit in which this oscillation terminates. This is because when the oscillation coefficient goes to 0, we will reach a critical point where the range of the particle's oscillation exceeds its allowed range.  In simpler words, since at the ISCO any little perturbation added to test particle will result the particle fall into plunge and never comes back,  if we added some perturbation before test particle reaches ISCO, this cross-limit scene must happen somewhere early than it reaches ISCO.

The way to work out this limit is simple. As we already worked out the amplitude for the test particle's oscillation, we can test that if it exist the `safe' region. The `safe' region is the maximum domain which the oscillation can happen. In effective potential it is simply the distance from the innermost stable circular orbit to innermost unstable circular orbit. As we can show that any particle crosses innermost unstable circular orbit will keep falling into Black Hole. 
\begin{figure}
\centering
\includegraphics[width=0.75\linewidth]{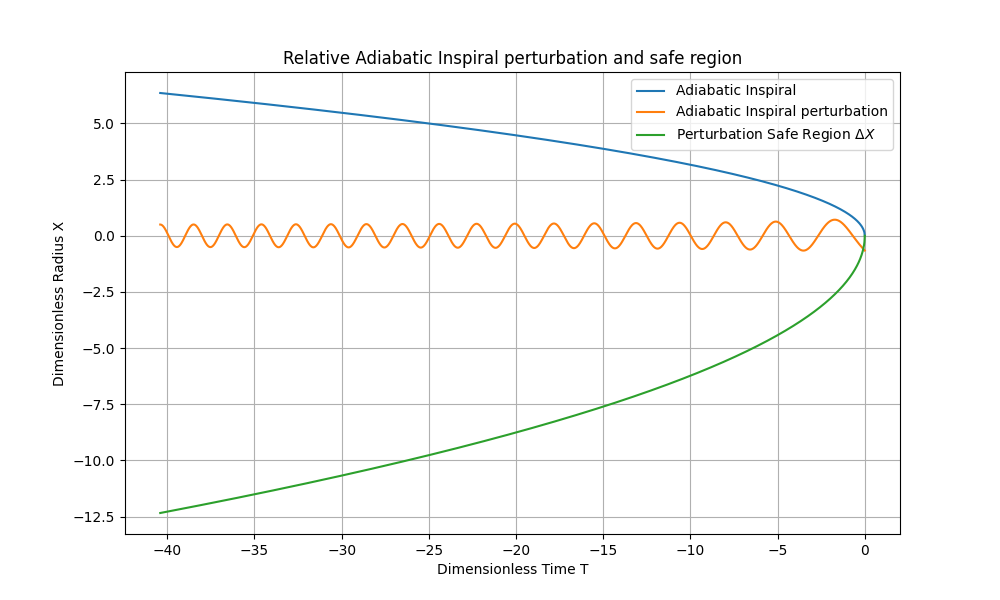}
\caption{\label{fig: aip3}Adiabatic Inspiral Relative Perturbation with Boundary. The blue line is the plot of Adiabatic inpiral $X(T)$, the orange line is the plot of adiabatic inspiral perturbation relative to Adiabatic Inspiral $\mathcal{X}(T)$, with $\tilde{a}=0$ (Schwarzschild), $\eta=10^{-5}$ and initial perturbation set as $\Delta X_0=0.5,\Delta v_0=0$ at $T=-40.4$  . The plot is generate using Fourth-order Runge-Kutta method with python, see Appendix  \ref{appendix: AIP Numerical Approximation}. The green line is  the perturbation safe region $\Delta X_{safe}$}
\includegraphics[width=0.75\linewidth]{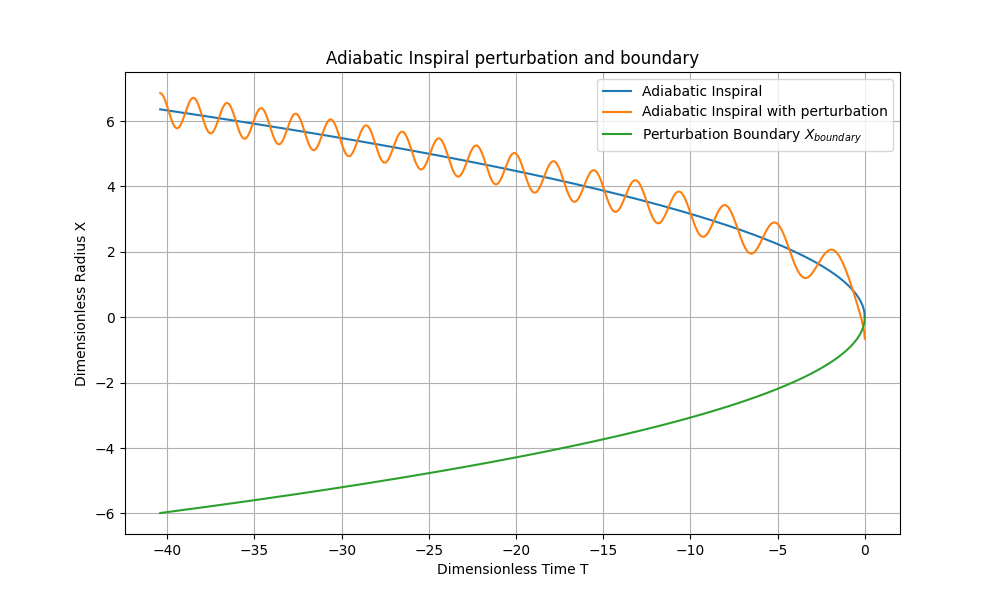}
\caption{\label{fig: aip4}Adiabatic Inspiral Perturbation with Boundary. This is the same plot of previous one but added perturbation onto the adiabatic inspiral. The orange  line is perturbed adiabatic inspiral solution $X_p(T)$ and green line is Perturbation Boundary $X_{boundary}$}
\end{figure}
As we also know that as test particle falls near ISCO the two circular orbit will eventually collapse at ISCO and effective potential forms a saddle point at ISCO, thus the safe region must shrinks. Using the formulas we obtains for the solution previously, the safe region is therefore: 
\begin{align*}
    \Delta \tilde{r}_{safe}=\tilde{r}_{c+}-\tilde{r}_{c-}=\sqrt{(\tilde{L}_s^2-\tilde{a}^2\tilde{E}_s^2+\tilde{a}^2)^2-12(\tilde{L}_s-\tilde{a}\tilde{E}_s)^2}.
\end{align*}
And the boundary radius is simply 
\begin{align*}
     \tilde{r}_{boundary}=\tilde{r}_s-\Delta \tilde{r}_{safe}.
\end{align*}
Thus whenever the test particle reaches critical radius $\tilde{r}_{boundary}$, it will fall into black hole and enter plunge regime immediately. 

Again, in-order to simplify this and use Ori-Thorne dimensionless notation, we consider the Schwarzschild case $\tilde{a}=0$. Then the expression reduces straight away to 
\begin{align*}
    \Delta \tilde{r}_{safe}=\frac{\tilde{r}_s(\tilde{r}_s-6)}{\left(\tilde{r}_s-3\right)}.
\end{align*}
For $\tilde{r}_s\geqslant\tilde{r}_{isco}$, which is trivial. And it is clear that at $\tilde{r}_s=\tilde{r}_{isco}=6$ , $\Delta \tilde{r}_{safe}=0$ meaning there is no allowed oscillation region, or equivalently that any little perturbation will terminate the stability. Now the boundary radius is:
\begin{align*}
    \tilde{r}_{boundary}=\tilde{r}_s-\Delta \tilde{r}_{safe}=\frac{3\tilde{r}_s}{\tilde{r}_s-3}.
\end{align*}

Now substitute in Ori-Thorne dimensionless notation we have:
\begin{align*}
    X_{boundary}&=\frac{\tilde{r}_{boundary}-\tilde{r}_{isco}}{R_0\eta^\frac{2}{5}}\\&=\frac{-3(R_0\eta^\frac{2}{5}X+6)+18}{(R_0\eta^\frac{2}{5})((R_0\eta^\frac{2}{5})X+3)}.
\end{align*}

For consistency with previous result, set $\eta:=10^{-5}$ and obtains:
\begin{align*}
    X_{boundary}&\simeq -\frac{0.088X}{0.00086X+0.088},\\
    \Delta X_{safe}&\simeq \frac{0.02925X^2+6X}{0.02925X+3}.
\end{align*}
As we can see in the plot \ref{fig: aip4}, the boundary is indeed shrinking and approaching to ISCO.

\subsection{Perturbation induced plunge phase}

As we previous discussed, as the perturbation boundary limits the maximum perturbation range that the test particle can oscillate. Once the oscillation goes over boundary, there will be no more oscillations. And the test particle will enter plunge phase and fall toward into Kerr Black Hole. In this chapter we shall derive this process.

The critical point where the Adiabatic Inspiral with Perturbation terminate, is at $X_p(T_{critical})=X_{boundary}(T_{critical})$. And the initial velocity is therefore:
\begin{align*}
    \frac{\text{d}X_p}{\text{d}T}|_{T_{critical}}=\frac{\text{d}X}{\text{d}T}|_{T_{critical}}+\frac{\text{d}\mathcal{X}}{\text{d}T}|_{T_{critical}}\\
    =-\frac{1}{2\sqrt{-T_{critical}}}+\frac{\text{d}\mathcal{X}}{\text{d}T}|_{T_{critical}}.
\end{align*}
The following process is simply plunge regime that we calculated above. Here we will continue to use Ori-Thorne Dimensionless notation and again assume that $\eta=10^{-5},\tilde{a}=0$, and the initial perturbation as in Appendix \ref{appendix: AIP Numerical Approximation}. Using numerical approximation we can obtain$$T_{critical}\simeq-0.0928,\frac{\text{d}\mathcal{X}}{\text{d}T}|_{T_{critical}}\simeq-0.8408.$$
Thus:
$$\frac{\text{d}X_p}{\text{d}T}|_{T_{critical}}\simeq-2.48.2$$
And for plunge regime, $X_{plunge}(T)=\frac{-\mathcal{x}}{(T_{plunge}-T)^2}$ cf. \cite{Ori_2000}. Thus using continuity and smoothness as boundary conditions, 
\begin{align*}
X_{plunge}(T_{critical})=X_p(T_{critical})=-0.3037,\\
  \frac{\text{d}X_{plunge}}{\text{d}T}|_{T_{critical}}=\frac{\text{d}X_p}{\text{d}T}|_{T_{critical}}\simeq-2.482,\\
  \Longrightarrow x\simeq0.01819,T_{plunge}\simeq0.1519.
\end{align*}
And our plunge phase equation is therefore:
\begin{align*}
    X_{plunge}(T)=\frac{-0.01819}{(0.1519-T)^2}.
\end{align*}

\begin{figure}
\centering
\includegraphics[width=0.75\linewidth]{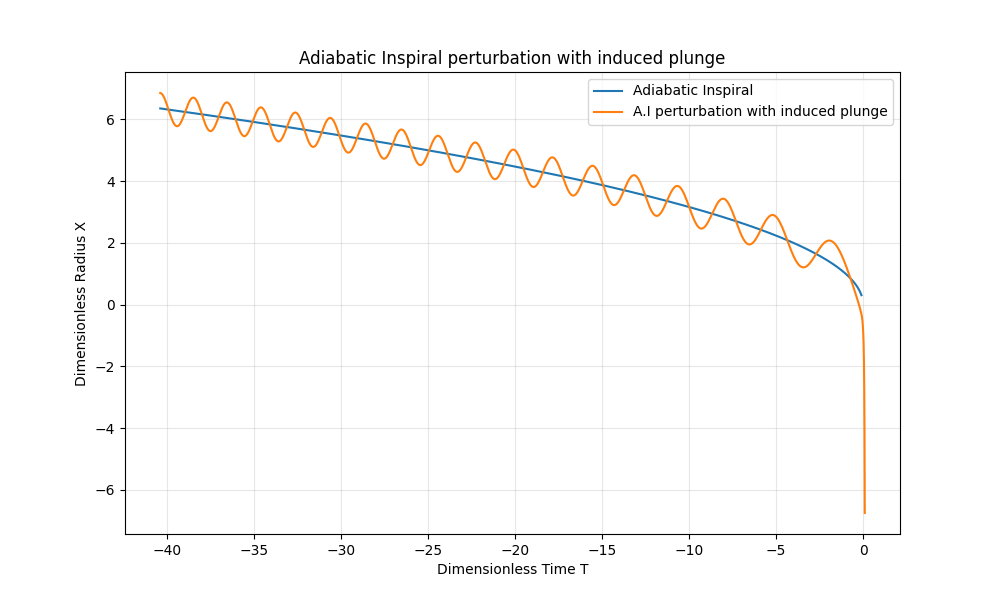}
\caption{\label{fig: aip5}Adiabatic Inspiral perturbation with induced plunge.  The orange line is adiabatic inspiral perturbation and its induced plunge phase.}
\end{figure}
The perturbation induced plunge suggest that the observation of plunge regime could be very distinct even for test particles following almost identical adiabatic inspiral, especially for almost-identical behaviour around Most Stable Circular Orbit. The initial boundary condition dominate the behaviour of plunge. Moreover, the plunges happens earlier than classical plunge, due to the fact that particle's oscillation will most likely take particle enters plunge before it actually reaches ISCO following adiabatic inspiral.

\appendix
\section{Appendix Ori-Thorne Numerical Approximation}
\label{appendix: OT Numerical Approximation}
For second-order ODE, we can use Euler's method to get a sequence of numerical approximate points. 
\begin{align*}
        \frac{\text{d}^2X}{\text{d}T^2} = -X^2 - T 
\end{align*}
Using Euler's method we can establish:
\begin{align*}
    T[i+1]&=T[i]+\Delta T, X[i+1]=X[i]+X'[i]\Delta T\\
    X'[i+1]&=X'[i]+X''[i]\Delta T=X'[i]+(-X[i]^2-T[i])\Delta T
\end{align*}
We then start by looking for the initial values and the steps.
\begin{align*}
    T[0],X[0],X'[0],\Delta T
\end{align*}
Once these values are obtained one can have the numerical solution of the ODE

To be consistent with our previous plot, plot \ref{fig:Figure3}, we would like to go  back to $\tilde{r},\tilde{\tau}$ coordinates. Similarly we use the Ori-Thorne ODE:
\begin{align*}
 \frac{\text{d}^2\tilde{r}}{\text{d}\tilde{\tau}^2} = -\alpha (\tilde{r}-\tilde{r}_\text{isco})^2 - \kappa\eta\beta\cdot\tilde{\tau} 
\end{align*}
and Euler's method:
\begin{align*}
        r[i+1]&=t[i]+\Delta t, r[i+1]=r[i]+r'[i]\Delta t\\
    r'[i+1]&=r'[i]+r''[i]\Delta t=r'[i]+(-\alpha(r[i]-\tilde{r}_{isco})^2-\kappa\eta\beta t[i])\Delta t
\end{align*}
And use the constants:
\begin{align*}
    \eta = 0.01, \tilde{a}=0.2
\end{align*}
Therefore we would have:

Using \ref{eq:r_isco solution SSCD} $\Longrightarrow$ $\tilde{r}_{isco}=5.32944329643$
\ref{eq: coordinat angular velocity SSCD} $\Longrightarrow$ $\tilde{\Omega}_{isco}=0.0799786762958$
And thus we can have:
\begin{align*}
    &\alpha =  \frac{1}{4} \frac{\partial^3 V_{\text{eff}}}{\partial \tilde{r}^3}(0, \tilde{r}_{\text{isco}})=0.00123957445205 \\
    &\beta =-\frac{1}{2}(\frac{\partial^2 V_{\text{eff}}}{\partial \tilde{E} \partial \tilde{r}}(0,\tilde{r}_{\text{isco}})\tilde{\Omega}_{isco}+\frac{\partial^2 V_{\text{eff}}}{\partial \tilde{L} \partial \tilde{r}}(0, \tilde{r}_{\text{isco}}))=0.0205677526263\\
\end{align*}
From Table I \cite{Ori_2000}: $\dot{\varepsilon}=1.114$, substitute these in \ref{eq:OT Transition Kappa} gives:
\begin{align*}
    \kappa=\frac{32}{5}\:\tilde{\Omega}_{isco}^\frac{7}{3}\dot{\varepsilon}\frac{1+\frac{\tilde{a}}{\tilde{r}^{\frac{3}{2}}_{isco}}}{\sqrt{1-\frac{3}{\tilde{r}_{isco}}+\frac{2\tilde{a}}{{\tilde{r}^\frac{3}{2}}_{isco}}}}=0.029139149184
\end{align*}
We now have all our constants determined. The next step is to identify the initial conditions. The transition regime is ideally applicable only near the ISCO radius. However, as discussed in the sections on adiabatic and plunge regimes, both approximations fail when the radius approaches the ISCO. This presents a significant challenge in justifying a reasonable starting point for the numerical solution of the transition. Near the ISCO radius, neither the adiabatic nor the plunge regimes provide an accurate initial point. Moreover, far from the ISCO, the transition regime itself becomes unreliable, further complicating the identification of an accurate initial condition. 

Here we will choose an initial point that locates between transition and adiabatic, by keep testing if that initial points works. The result is that for $\tilde{r}\gtrsim5.4$ under this setting the transition approximation fails. Thus we can try to set $r[0]=5.4$ 

Substitute that in \ref{eq: Adiabatic} gives $t[0]=-10.3855852165$, and differentiate both sides of \ref{eq: Adiabatic} we have:

\begin{align*}    \text{d}\tilde{\tau}/\text{d}\tilde{r}|_{\text{start}}=-295.57524287\:\Longrightarrow r'[0]=\text{d}\tilde{r}/\text{d}\tilde{\tau}=-0.00338323328534
\end{align*}
And we select $\Delta t=0.1$ and use the python code to generate points up until $\tilde{r}\leq0$:
\begin{verbatim}
r_i=r_0
t_i=t_0
r_prime_i=r_prime_0
for i in range(2000):
    new_t_i=t_i+dt
    new_r_i=r_i+r_prime_i*dt
    new_r_prime_i=r_prime_i+(-alpha*(r_i-r_isco)*(r_i-r_isco)-kappa*eta*beta*abs(t_i))*dt
    if new_r_i<0:
        break
    print(r_i,`,',t_i)
    r_i=new_r_i
    t_i=new_t_i
    r_prime_i=new_r_prime_i
\end{verbatim}
Here we reverse the order of $\tilde{r}$ and $\tilde{\tau}$ to be consistent with the previous plots. The code will yield$~1300$ pairs of results.

\section{Appendix AIP Numerical Approximation}
\label{appendix: AIP Numerical Approximation}
The python code for this Numerical Approximation is:
\begin{verbatim}
import numpy as np
import matplotlib.pyplot as plt

#Define ODE
def f(X, v, T):
    coeff = np.sqrt(-T) / (0.0007716 * (0.0295*np.sqrt(-T) + 6)**3 * (0.0295*np.sqrt(-T) + 3))
    return v, -coeff*X

# Fourth-order Runge-Kutta method  (RK4)
def rk4_step(X, v, T, dT):
    k1X, k1v = f(X, v, T)
    k2X, k2v = f(X + 0.5*dT*k1X, v + 0.5*dT*k1v, T + 0.5*dT)
    k3X, k3v = f(X + 0.5*dT*k2X, v + 0.5*dT*k2v, T + 0.5*dT)
    k4X, k4v = f(X + dT*k3X, v + dT*k3v, T + dT)
    
    X_next = X + (dT/6)*(k1X + 2*k2X + 2*k3X + k4X)
    v_next = v + (dT/6)*(k1v + 2*k2v + 2*k3v + k4v)
    return X_next, v_next

# Initial Perturbation Condition
X0 = 0.5  # Initial Perturbation Position
v0 = 0  
T0 = -40.4 #Initial Perturbation Time
Tf = 0  
N = 2000  # Timestep
dT = (Tf - T0) / N  

T = np.linspace(T0, Tf, N)
X = np.empty(N)
v = np.empty(N)
X[0], v[0] = X0, v0

for i in range(1, N):
    X[i], v[i] = rk4_step(X[i-1], v[i-1], T[i-1], dT)

plt.figure(figsize=(10, 6))
plt.plot(T, np.sqrt(-T), label=`Adiabatic Inspiral')
plt.plot(T, X+np.sqrt(-T), label=`Adiabatic Inspiral with perturbation')
plt.plot(T, -(0.08775*np.sqrt(-T)) / (0.0008555625*np.sqrt(-T)+0.08775),
    label=`Perturbation Boundary $X_{boundary}$')
#plt.plot(T, v, label='Velocity (dX/dT)')
plt.xlabel('Dimensionless Time T')
plt.ylabel('Dimensionless Radius X')
plt.legend()
plt.title(`Adiabatic Inspiral perturbation and boundary')
plt.grid(True)
plt.show()

\end{verbatim}

\bibliographystyle{alpha}
\bibliography{references}

\end{document}